\pdfoutput=1
\documentclass[iop,revtex4,letterpaper]{emulateapj}
\usepackage{lscape}
\usepackage{longtable, subfigure}
\usepackage{amsmath}
\usepackage{lmodern}
\usepackage[T1]{fontenc}
\usepackage{natbib}
\usepackage{graphicx}
\usepackage[space]{grffile}
\usepackage{latexsym}
\usepackage{multirow}
\usepackage{amsfonts,amsmath,amssymb}
\usepackage{url}
\usepackage[utf8]{inputenc}
\usepackage{fancyref}
\usepackage{hyperref}
\usepackage{dcolumn}
\usepackage{verbatim}

\hypersetup{colorlinks=false,pdfborder={0 0 0},}
\slugcomment{DRAFT \today}
\shorttitle{Low-Mass Pleiades Periods}
\shortauthors{Covey et al.}

\begin{document}


\newcommand{\lapprox }{{\lower0.8ex\hbox{$\buildrel <\over\sim$}}}
\newcommand{\gapprox }{{\lower0.8ex\hbox{$\buildrel >\over\sim$}}}
\newcommand{\Msun}{\ifmmode {M_{\odot}}\else${M_{\odot}}$\fi}
\newcommand{\Rsun}{\ifmmode {R_{\odot}}\else${R_{\odot}}$\fi}
\def\asec{\ifmmode^{\prime\prime}\else$^{\prime\prime}$\fi}
\def\ctss{$10^{-2}$ counts s$^{-1}$}
\def\ergs{$\mathrm{erg}$~$\mathrm{s}^{-1}$}
\def\ergcms{$\mathrm{erg}$~$\mathrm{cm}^{-2}$~$\mathrm{s}^{-1}$}
\def\ROSAT{\it ROSAT}
\def\CHANDRA{\textit{Chandra}}
\def\LL{$L_{\mathrm{X}}/L_{\mathrm{bol}}$}
\def\Prot{$P_{\mathrm{rot}}$}
\def\Pmem{$P_{\mathrm{mem}}$}

\title{Why are Rapidly Rotating M Dwarfs in the Pleiades so (Infra)red? \\ 
New Period Measurements Confirm Rotation-Dependent Color Offsets From the Cluster Sequence}

\author{Kevin~R.~Covey\altaffilmark{1,2}, 
Marcel~A.~Ag\"ueros\altaffilmark{3}, 
Nicholas~M.~Law\altaffilmark{4}, 
Jiyu Liu\altaffilmark{3}, 
Aida Ahmadi\altaffilmark{5},
Russ Laher\altaffilmark{6},
David Levitan\altaffilmark{7},
Branimir Sesar\altaffilmark{8},
Jason Surace\altaffilmark{6}}

\altaffiltext{1}{Department of Physics \& Astronomy, Western Washington University, Bellingham WA 98225-9164 USA \email{kevin.covey@wwu.edu}}
\altaffiltext{2}{Lowell Observatory, 1400 W.~Mars Hill Rd., Flagstaff AZ 86001 USA}
\altaffiltext{3}{Department of Astronomy, Columbia University, 550 West 120th Street, New York, NY 10027, USA}
\altaffiltext{4}{Department of Physics \& Astronomy, University of North Carolina, Chapel Hill, NC 27599-3255, USA}
\altaffiltext{5}{Max Planck Institute for Radioastronomy, Auf dem H\"ugel 69, 53121 Bonn, Germany}
\altaffiltext{6}{Spitzer Science Center, California Institute of Technology, Pasadena, CA 91125, USA}
\altaffiltext{7}{Division of Physics, Mathematics, and Astronomy, California Institute of Technology, Pasadena, CA 91125, USA}
\altaffiltext{8}{Max Planck Institute for Astronomy, K\"onigstuhl 17, D-69117 Heidelberg, Germany}

\begin{abstract}
Stellar rotation periods (\Prot) measured in open clusters have proved to be extremely useful for studying stars' angular momentum content and rotationally driven magnetic activity, which are both age- and mass-dependent processes. While \Prot\ measurements have been obtained for hundreds of solar-mass members of the Pleiades, measurements exist for only a few low-mass ($<$0.5~\Msun) members of this key laboratory for stellar evolution theory. To fill this gap, we report \Prot\ for 132 low-mass Pleiades members (including nearly 100 with $M \leq 0.45$ \Msun), measured from photometric monitoring of the cluster conducted by the Palomar Transient Factory in late 2011 and early 2012. These periods extend the portrait of stellar rotation at 125 Myr to the lowest-mass stars and re-establish the Pleiades as a key benchmark for models of the transport and evolution of stellar angular momentum. Combining our new \Prot\ with precise $BVIJHK$ photometry reported by Stauffer et al.~and Kamai et al., we investigate known anomalies in the photometric properties of K and M Pleiades members. We confirm the correlation detected by Kamai et al.~between a star's \Prot\ and position relative to the main sequence in the cluster's color-magnitude diagram. We find that rapid rotators have redder $(V-K)$ colors than slower rotators at the same $V$, indicating that rapid and slow rotators have different binary frequencies and/or photospheric properties. We find no difference in the photometric amplitudes of rapid and slow rotators, indicating that asymmetries in the longitudinal distribution of starspots do not scale grossly with rotation rate.
\end{abstract}


\section{Introduction}
Studies of stellar rotation in the Pleiades go back decades; indeed, Pleiads are included in the seminal analysis by \citet{skumanich72} of the evolution of rotation and activity in Sun-like stars. Spectroscopic $v_{rot}$~sin~$i$ measurements were quickly obtained for large numbers of cluster members \citep[e.g.,][]{Stauffer1987}, but inclination-independent measurements of rotation periods (\Prot) took longer to accumulate, reflecting the observationally intensive nature required of monitoring programs conducted with photometers or small field-of-view imagers  \citep[e.g.,][]{vanLeeuwen1987, Prosser1993}. This changed dramatically with the HATNet exoplanet-search program, which monitored $\approx$100 square degrees including much of the cluster and produced periods for 368 Pleiads \citep{Hartman2010}. However, while these authors achieved a 93\% period-detection efficiency for 0.7--1.0~\Msun\ members, this efficiency dropped sharply with mass, and the {number} of M-dwarf {Pleiads} with measured \Prot\ remains very small \citep[e.g., the few studied by][]{Terndrup1999,Scholz2004}. 

Measurements of \Prot\ for single-age open-cluster members spanning a range of masses {are a valuable way to test} models of stellar angular-momentum evolution. These models strive to reproduce the dependence of \Prot\ on mass within a cluster and the evolution of \Prot\ for stars in a narrow mass range (e.g., the age-rotation relation for 0.8--1.1 \Msun\ stars). The recent review by \citet{Bouvier2014} provides a fuller overview of the increasingly sophisticated theoretical descriptions of processes responsible for angular-momentum loss and/or transfer. Broadly speaking, the components of such models are: a) the initial distribution of angular-momentum states \citep[e.g.,][]{Joos2012}; b) the mechanism and efficiency of angular-momentum loss during the protostellar/accretion phase \citep[][]{Matt2005, Romanova2009}; c) the efficiency and timescale of angular-momentum transport in the stellar interior \citep[e.g.,][]{Denissenkov2010}; and d) the efficiency of angular-momentum loss from the zero-age main sequence {(ZAMS) onward} \citep[e.g.,][]{Matt2015, Gallet2015}.  

The rotation rates of Pleiades members have often been particularly useful in this context, as these can be used to characterize the angular-momentum content of stars that have just arrived on the {ZAMS} \citep[e.g.,][]{Bouvier1997, Sills2000}. However, the lack of \Prot\ measurements for the lowest-mass Pleiads has led authors seeking rotation periods for Pleiades-age stars to utilize a {pseudo-Pleiades sample of} \Prot\ measured for high-mass members of M35 \citep{Meibom2009} {and} periods measured for low-mass members of NGC 2516 \citep{Irwin2007}. While combining distinct populations is sufficient for first-order investigations of angular-momentum evolution, {testing second-order} effects due to metallicity, cluster environment, etc., requires that we obtain full mass sequences in many cluster environments, particularly one as well studied as the Pleiades.

The spot signatures from which \Prot\ values are inferred also provide important clues as to the behavior of active photospheres and their influence on stellar properties. Early studies identified systematic differences between the photometric colors of members of the Pleiades and older open clusters such as the Hyades and Praesepe \citep{Stauffer1984a}, and between the colors of rapidly and slowly rotating Pleiads \citep[e.g., figure 5 in][]{Stauffer1984b}. Subsequent studies have documented the anomalous morphology of the Pleiades cluster sequence \citep[e.g.,][]{Bell2012} and provided further evidence of connection between this morphology and rotation rates \citep{Stauffer2003,Kamai2014}. Magnetically confined starspots provide a likely physical mechanism for linking a star's rotation rate and photospheric properties: rotationally driven magnetic dynamos produce large cool spots on the photospheres of rapidly rotating stars, which simple two-component spot models predict produce photometric anomalies similar to those which are observed \citep[e.g.,][]{Stauffer1986,Jackson2013}.

Uncertainties in empirical estimates of stellar parameters derived from these anomalous photometric properties could produce systematic errors in the ages, masses, and radii of spotted stars, with important ramifications for the inferred age scale for young (1-200 Myr) stars and clusters. In addition to these observational effects, starspots can significantly affect the energy transport, temperature structure, and lithium-depletion efficiency in pre-main sequence stellar interiors \citep[e.g.,][]{Jackson2014b,Jackson2014a,Somers2015a,Somers2015b}, introducing additional systematic uncertainties into the parameters inferred by comparison to non-spotted stellar evolutionary models. 

To improve empirical constraints on the rotational evolution and photospheric properties of low-mass stars, we carried out a sensitive, wide-field multi-epoch monitoring campaign targeting the lowest-mass members of the 125-Myr-old Pleiades cluster. We begin in Section~\ref{sec:catalog} by using the literature to assemble a catalog of $>$2300 Pleiades members with reliable photometry spanning $K = 5-19$ mag. We then use the absolute $K$ magnitudes of these stars to derive their masses, and identify candidate binaries in our sample based on their position in a $V$ versus $(V-K)$ color-magnitude diagram (CMD). In Section~\ref{sec:periodMeasure}, we describe our Palomar Transient Factory \citep[PTF;][]{Law2009,rau2009} observations of four fields in the Pleiades that provide PTF light curves for 809 candidate members of the cluster. We then present our period-finding pipeline, along with a number of tests we developed to establish the reliability of our {132} \Prot\ measurements for Pleiads ranging from 0.18 to 0.65 \Msun. These measurements include the first \Prot\ values reported for $\approx$100 low-mass ($M\leq0.45$ \Msun) cluster members. We discuss our results in Section~\ref{sec:discuss} {and conclude} in Section~\ref{sec:concl}. {Finally, we include in the Appendix 119 variable field stars identified in our Pleiades observations.}

\section{Catalog Assembly}\label{sec:catalog}
\subsection{Membership Catalog}\label{sec:PleiadesMembers}
We {collate} existing catalogs of confirmed and candidate Pleiads to characterize the subset for which PTF collected densely sampled light curves. As the basis of this catalog, we adopt the \citet{Stauffer2007} list of Pleiades members. This compilation includes 1416 candidate members identified by others over more than 80 years \citep[e.g., by][]{Trumpler1921,Artyukhina1969,vanLeeuwen1986,Deacon2004}.

To produce a uniform set of astrometric and photometric measurements for these stars, \citet{Stauffer2007} identified counterparts for brighter ($J < 11$ mag) objects in the 2MASS All-Sky Point Source Catalog \citep{cutri03} and for fainter objects in the deeper ``6x'' catalog that 2MASS produced by observing a 3 degree $\times$ 2 degree field centered on the Pleiades with exposure times six times longer than those used in the primary survey. 

We supplement the \citet{Stauffer2007} catalog with additional candidate Pleiades members identified and assembled by \citet{Lodieu2012}, \citet{Bouy2015} and \citet{Hartman2010}. Using photometry and astrometry from the UKIRT Infrared Deep Sky Survey (UKIDSS), \citet{Lodieu2012} identified 1076 stars within 5$^{\circ}$ of the Pleiades's center for which they calculate a membership probability (\Pmem) $>$50\%. Using color, magnitude, and proper motion cuts, \citet{Lodieu2012} also selected an overlapping, but non-probabilistic set of 1147 candidate low-mass Pleiads.  To test the fidelity and completeness of these UKIDSS-selected samples, \citet{Lodieu2012} compiled a list of candidate Pleiades members identified over the past two decades \citep[e.g., by][]{Hambly1993,Festin1998,Pinfield2000,Moraux2003,Bihain2006}.

Using a 3\arcsec\ radius, we {match} these lists and {identify} 506 previously reported candidate Pleiades members that are not in the \citet{Stauffer2007} list, and 466 additional candidate members reported by \citet{Lodieu2012}.\footnote{The 1314 candidate Pleiades members in appendix C of \citet{Lodieu2012} {are} mislabeled. {This} is the complete list of candidate cluster members, not just newly identified members.} 

We also incorporate candidate Pleiades members identified by \citet{Bouy2015} from their updated DANCe-Pleiades catalog, originally developed by \citet{Bouy2013} and \citet{Sarro2014}. The DANCe-Pleiades catalog {includes} photometry from several wide-field optical {and infrared} surveys, 
as well as dedicated imaging programs carried out with imagers on telescopes in single-user mode. 
{Combining these data} along with new Y-band observations from the {\it William Herschel Telescope}, \citet{Bouy2015} produce two catalogs with calculated Pleiades membership probabilities, for sources with and without Tycho-2 photometry {and} astrometry.  Tycho-2 sources are sufficiently bright that they saturate a standard {PTF} exposure, so we include only the 2010 candidates without Tycho-2 counterparts for which \citet{Bouy2015} calculate \Pmem {$> 75$\%}. Using a 3\arcsec\ matching radius, we {find} 1606 DANCe-Pleiades candidates with counterparts in the merged \citet{Stauffer2007}/\citet{Lodieu2012} catalog, and incorporate the remaining 404 unmatched DANCe-Pleiades candidates into our catalog. 

Finally, we include {seven} candidate Pleiads for which \citet{Hartman2010} measured \Prot, but that lack counterparts within 3\arcsec\ in our combined \citet{Stauffer2007}/\citet{Lodieu2012}/\citet{Bouy2015} catalog. The composition of this merged catalog of Pleiades members and candidates is summarized in Table~\ref{tab:membership}. {We} have 2799 potential Pleiads ranging from $K = 5-19$ mag. 

\begin{figure}[t!] 
\centerline{\includegraphics[angle=0, width=\columnwidth]{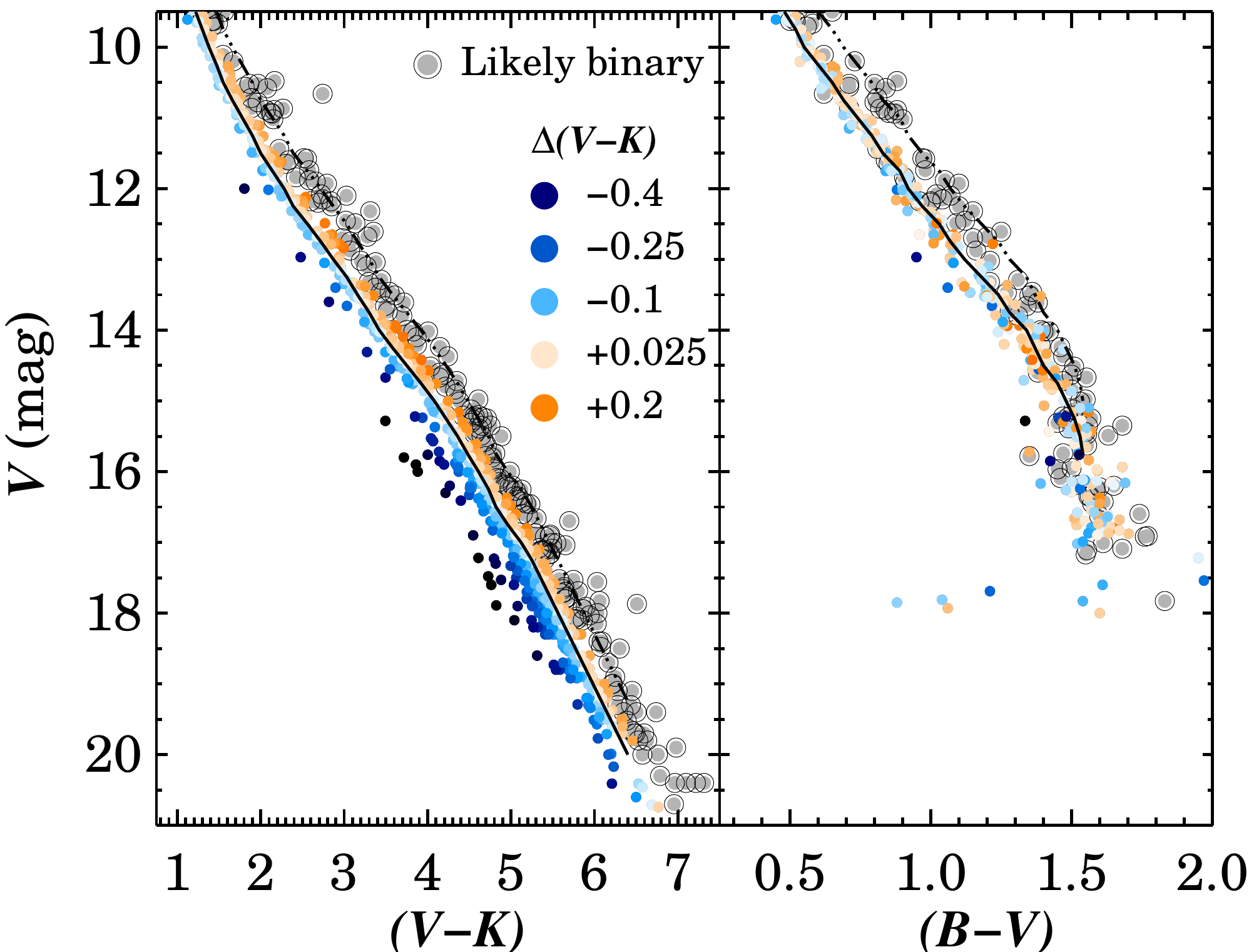}}
\caption{$V$ versus $(V-K)$ \textit{(left)} and versus $(B-V)$ \textit{(right)} CMDs for Pleiades members. Stars are color-coded according to their location relative to the semi-empirical $V$ versus $(V-K)$ cluster sequence of \citet{Kamai2014}, and which we extend to $V=20$ and $(V-K) = 6.4$ (solid line in both panels). Dash-dotted lines show the same sequence offset brightward by 0.75 mag to indicate the expected position of equal-mass binary systems. Likely binaries, selected as members with $V$ more than {0.375} mag brighter than the $V$ versus $(V-K)$ sequence, are indicated with gray circles. }
\label{fig:ColorOffsets}
\end{figure}

\subsection{Photometry}
Inferring reliable, self-consistent parameters for Pleiades members over a broad range in color, and thus mass, requires a comprehensive and homogeneous catalog of optical and near-infrared (NIR) photometry. At optical wavelengths, $R_{PTF}$ magnitudes are available for lower-mass Pleiads, but not for the higher-mass stars with \Prot\ reported by \citet{Hartman2010}.  Fortunately, the cluster's importance as a testbed for stellar evolution models has motivated the collection of $BVRI$ photometry for many of its members. 

We use the photometry compiled by \citet{Stauffer2007}, supplemented with the deep $BVI$ photometry {of} \citet{Kamai2014} for low-mass Pleiads. These two catalogs provide consistent $V$ magnitudes for $\approx$80\% (100/132) of the Pleiads for which we have \Prot\ from PTF light curves, and $\approx$88\% (436/496) of the full catalog of Pleiades members with measured \Prot. 

The resulting $V$ versus $(V-K)$ and $V$ versus $(B-V)$ CMDs for Pleiades members are shown in Figure~\ref{fig:ColorOffsets}. While the coverage of the full cluster population is not ideal, the larger photometric uncertainties in the transformed $R_{PTF}$ 
photometry would limit the analysis of color anomalies presented in Section~\ref{sec:color_rotation} much more than the modest reduction in sample size that results from adopting the $BVRI$ photometry. 

\begin{figure}[t!] 
\centerline{\includegraphics[angle=0,width=\columnwidth]{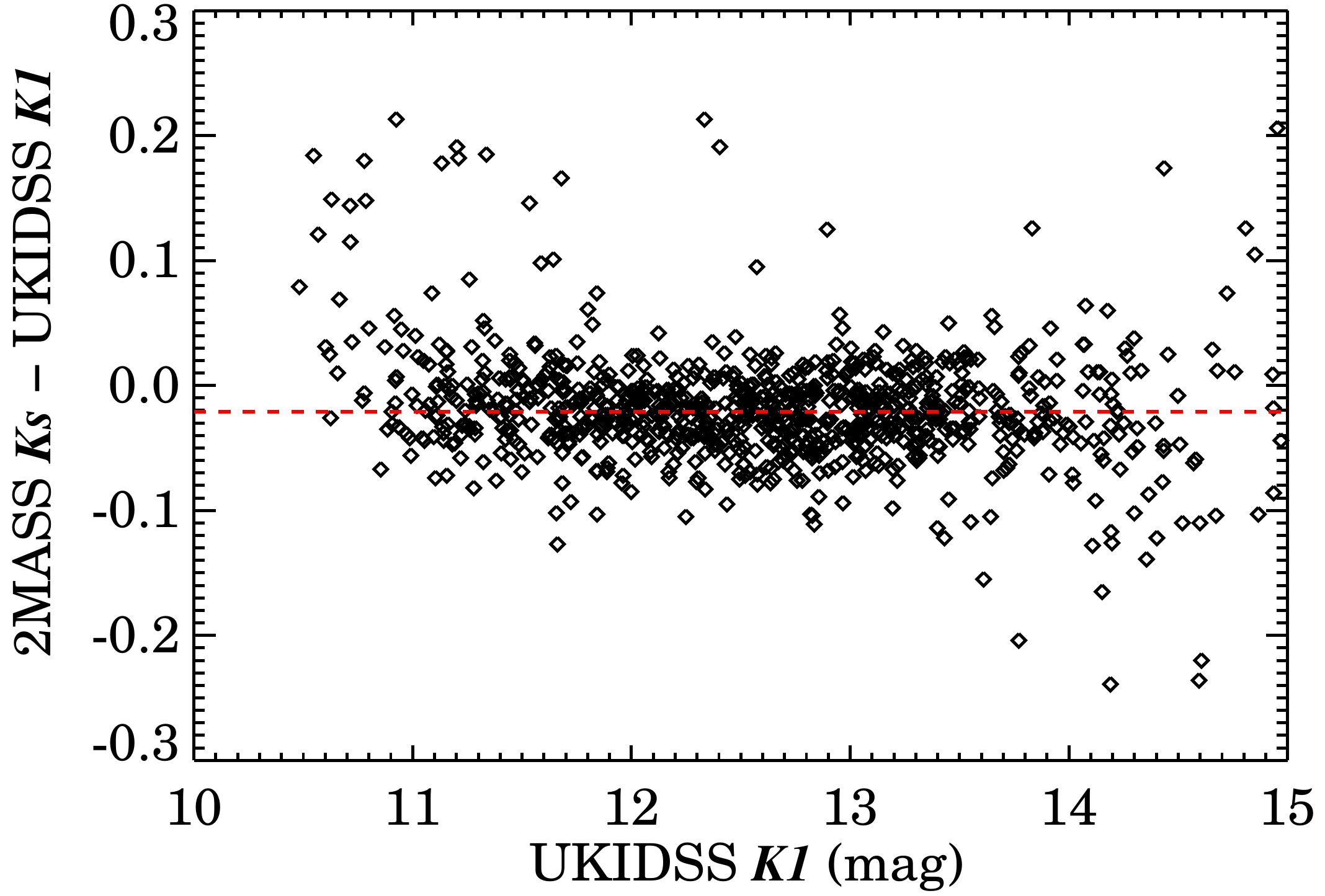}}
\caption{2MASS $K_s$ -- UKIDSS $K1$ residuals for sources in both surveys. The median offset is 0.02 mag; Malmquist-like effects produce small systematic biases for the very brightest and faintest sources detected in both surveys, but no large-scale trends with source magnitude are visible over the majority of the magnitude range in common to the two surveys. Tests of the agreement between the 2MASS and UKIDSS $J$ and $H$ show similarly small offsets (0.05 and 0.02 mag in the mean, respectively) and no large-scale trends as a function of color or magnitude.  } \label{fig:Kdiffs}
\end{figure}	

At NIR wavelengths, 2MASS and UKIDSS photometry are available for the full sample of Pleiades members. To remove potential systematic effects due to small differences between the photometric systems, we measured the offsets required to bring the UKIDSS $JHK$ onto the 2MASS photometric system within the central 1.3--1.5 magnitudes of the overlap range between the two surveys. Figure~\ref{fig:Kdiffs} shows the good agreement between the $K$ magnitudes measured for Pleiades members by 2MASS and UKIDSS, where systematic offsets can only be seen for the very brightest and faintest sources detected by both surveys. We found median 2MASS -- UKIDSS offsets of $-0.05$, $0.02$, and $-0.02$ mag in the $J$, $H$ and $K$ bands, respectively. {These offsets are consistent with the modest color terms and zero-point offsets between the WFCAM and 2MASS systems measured by \citet{hodgkin2009}.} 

\begin{figure*}[!t] 
\centerline{\includegraphics[angle=0,width=2\columnwidth]{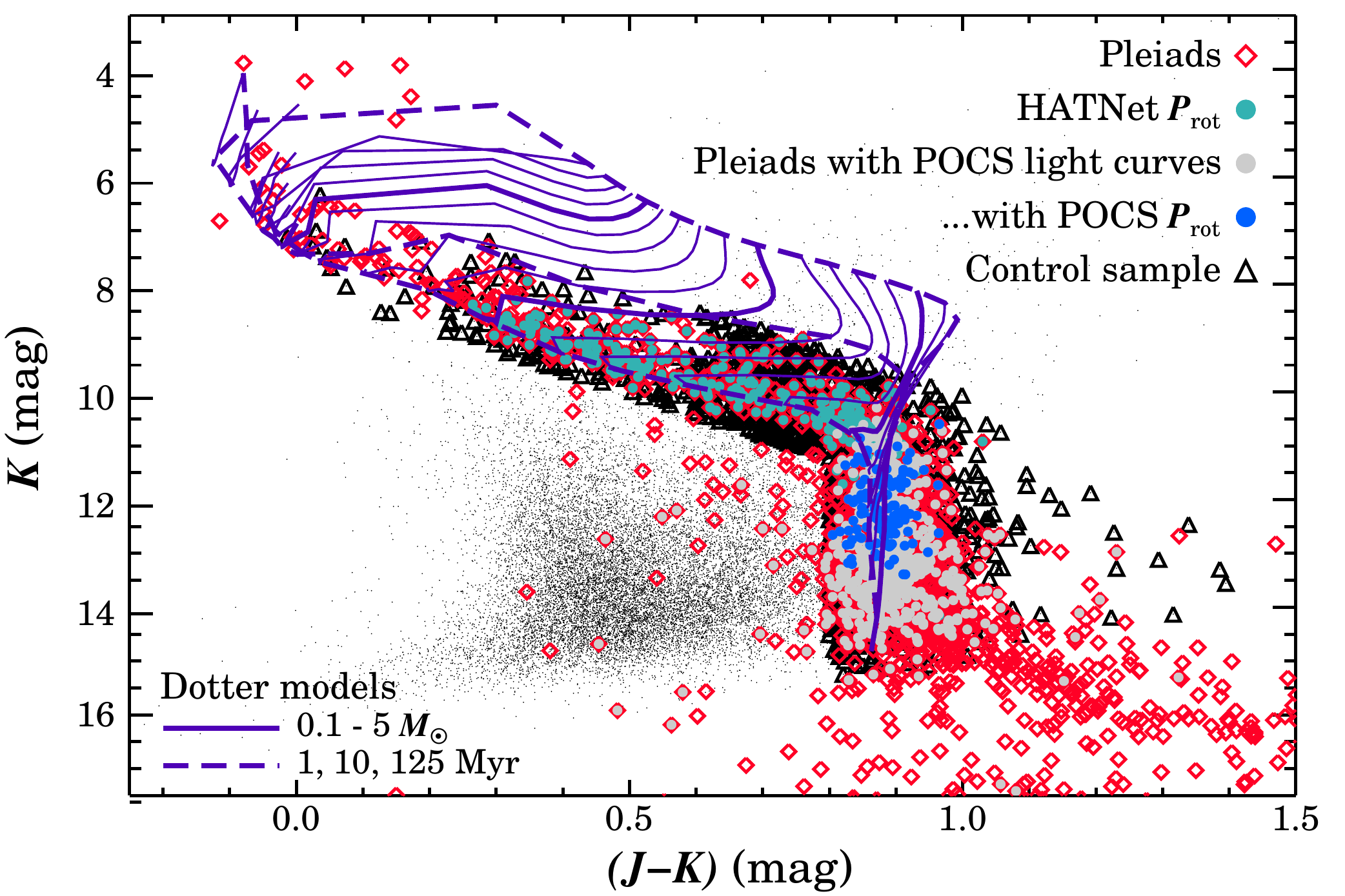}}
\caption{$(J-K)$ versus $K$ CMD for candidate Pleiades members and the control sample discussed in Section~\ref{sec:controlSample}; also shown for context are other sources in the 2MASS photometric catalog (black points) and solar-metallicity isochrones (1, 10, and 125 Myr; dashed lines) and mass tracks (0.1--5 \Msun; solid lines) computed for the 2MASS filters by A.~Dotter (priv.~communication). Despite combining photometry from the 2MASS and UKIDSS surveys, our catalog of candidate members produces a well-defined cluster sequence over more than 10 magnitudes in $K$ and a wide range of stellar masses ($0.07 < M < 2.5$ \Msun).} 

\label{fig:JKvsK_CMD}
\end{figure*}

We adopt 2MASS magnitudes for all sources with $K_s < $ 13 ($\approx$1.3 mag brighter than the 2MASS faint limit) or lacking UKIDSS magnitudes. {Only a few stars lacking 2MASS photometry are among our final sample of rotators (i.e., four of 132); for these sources, we adopt UKIDSS magnitudes after applying a simple zero-point correction to place them on the 2MASS system}. The $(J-K)$ versus $K$ CMD resulting from this combined photometric catalog is shown in Figure~\ref{fig:JKvsK_CMD}, and traces out a well-defined cluster sequence over more than 10 magnitudes in $K$.  

\begin{deluxetable*}{cccccc}[!h]
\tablewidth{0pt}
\tabletypesize{\scriptsize}
\tablecaption{Pleiades Members and Candidates}
\tablehead{
\colhead{} & \colhead{} & \colhead{Members or} & \colhead{Added} & \colhead{}  & \colhead{With} \\
\colhead{} & \colhead{$K$} & \colhead{Candidates} & \colhead{Members or} & \colhead{With PTF}  & \colhead{PTF} \\
\colhead{Catalog} & \colhead{(mag)} & \colhead{Cataloged} & \colhead{Candidates} & \colhead{Light Curves}  & \colhead{\Prot} 
}
\startdata
\multicolumn{1}{l}{\citet{Stauffer2007}} & 5--15 & 1416 & 1416 & 586 & 118 \\
\multicolumn{1}{l}{\citet{Lodieu2012}} & 11--19 & 2384 & 972 & 203 & 10 \\
\multicolumn{1}{l}{\citet{Bouy2015}} & 6.5--17.5 & 2010 & 404 & 29 & 4 \\
\multicolumn{1}{l}{\citet{Hartman2010}} & 9.5--12 & 383 & 7 & 0 & 0 \\
\hline 
\multicolumn{1}{l}{Control sample\tablenotemark{a}} & 7--15 & 2024 & \nodata & 424 & 3
\enddata
\tablenotetext{a}{See description in Section~\ref{sec:controlSample}.}
\label{tab:membership}
\end{deluxetable*}

\subsection{Stellar Parameters}\label{sec:stellar_parameters}
Even with the available high-quality photometry, estimating reliable stellar parameters for Pleiades-age stars is more difficult than it may appear. As noted earlier, \citet{Stauffer2003}, \citet{Bell2012} and \citet{Kamai2014} detected offsets between the empirical Pleiades CMD and that predicted by theory or empirically measured in older open clusters (i.e., Praesepe and the Hyades), an effect these authors attribute to the presence of cool starspots on the stellar photosphere. \citet{Stauffer2003} and \citet{Kamai2014} also reported correlations between each star's rotation rate ($v_{rot}$ sin $i$ and \Prot\, respectively) and its color/magnitude displacement, providing additional evidence that this offset is a signature of the temperature differences introduced by large starspots on the photospheres of the Pleiades's fastest rotating, and thus most magnetically active, low-mass members. We utilize our expanded sample of \Prot\ and homogeneous photometric data to examine this idea in Section~\ref{sec:color_rotation}; for now, we simply note that the presence of this offset complicates the assignment of stellar parameters based on a member's colors and magnitudes.

To make matters worse, the color-temperature relations predicted by various models and empirical calibrations \citep[e.g.,][]{Dotter2008,Boyajian2012,Pecaut2013} can disagree by as much as several hundred K, adding yet another systematic uncertainty when converting from observed colors to masses and temperatures. Luckily, these problems appear to be due to discrepancies at optical wavelengths: \citet{Bell2012} find that these offsets diminish to a negligible level at wavelengths longer than 2.2 $\mu$m, indicating that potential errors in photometric mass estimates can be minimized by inferring masses from a star's $K$ magnitude.

We therefore {convert} each Pleiades member's $K$ to an absolute $M_K$, adopting the distance modulus of $m-M=$ 5.67 measured by \citet{Melis2014} from VLBA parallaxes.\footnote{The distance implied by the parallax measurements in the revised {\it Hipparcos} catalog disagrees with that measured by \citet{Melis2014} as well as by other authors \citep[e.g.,][]{Pinsonneault2004,Soderblom2005} at the 
$\approx$5$\sigma$ level according to the formal errors in each study. This distance difference corresponds to a $\approx$10\% systematic uncertainty in the masses we infer, a relatively modest effect in the context of a sample which spans an order of magnitude in mass ($0.1 < M < 1.0$ \Msun).} We then {infer the} masses using the mass-$M_K$ relationship predicted by a 125 Myr, solar-metallicity ([Fe/H] = 0), non-$\alpha$-enhanced ([$\alpha$/Fe] = 0) Dartmouth isochrone \citep{Dotter2008}. We present these masses in Table~\ref{tab:Periods}, along with other relevant stellar parameters, for all stars for which we extract a robust \Prot. 
  
Finally, we {label} candidate binaries by analyzing the location of cluster members in the $V$ versus $(V-K)$ CMD. We {follow} \citet{Steele1995}, who demonstrated that synthetic binary systems form a second sequence in the $I$ versus $(I-K)$ CMD that lies 0.75 mag brightward of the single-star sequence. These authors also showed that the binary sequence included systems with a wide range of mass ratios: even the lowest mass secondaries emit enough {NIR} flux to shift a binary system significantly redward of the single-star sequence in an optical/{NIR} CMD, and thus into the elevated binary sequence at that redder color. Only when the mass ratio becomes extremely uneven ($M_{prim} / M_{sec}$ = 0.2/0.03 $\approx$ 7) does the secondary fail to contribute enough red flux to push the system well into the elevated binary sequence.

We indicate the location of this binary sequence in Figure~\ref{fig:ColorOffsets} by applying a 0.75 mag offset to the semi-empirical cluster sequence defined by \citet{Kamai2014}, which we extend here to $V=20$ and $(V-K)=6.4$. {We identify candidate binaries as those cluster members with $V$ within 0.375 mag of the binary sequence in Figure~\ref{fig:ColorOffsets}; this assumes a 0.375 mag spread in both the binary and single-star populations. We flag the candidate binaries in Table~\ref{tab:Periods}, and investigate the effects of this binary selection threshold further in Figure \ref{fig:signif_vs_vk} and Section \ref{sec:color_rotation}.} 

\section{Period Measurements}\label{sec:periodMeasure}
\subsection{Photometric Monitoring and Light Curve Construction}
We monitored the Pleiades using time allocated to two PTF Key Projects: the PTF/M-dwarfs survey \citep{Law2011, Law2012} and the PTF Open Cluster Survey \citep[POCS;][]{Agueros2011, Douglas2014}. PTF was a time-domain experiment using the robotic 48-inch Samuel Oschin (P48) telescope at Palomar Observatory, CA, and involved real-time data-reduction and transient-detection pipelines and a dedicated follow-up telescope. 

The PTF infrastructure is described in \citet{Law2009}; we focus here on the components associated with the P48, which we used to conduct our monitoring campaign. The P48 is equipped with the CFH12K mosaic camera, which was significantly modified to optimize its performance on this telescope. The camera has 11 working CCDs, which cover a 7.26 square degree field-of-view with 92 megapixels at 1\arcsec~sampling \citep{Rahmer2008}. Under typical observing conditions (1$\farcs$1~seeing), it delivers 2\arcsec~full-width half-maximum images that reach a 5$\sigma$ limiting $R \approx21$~mag in 60~s \citep{Law2010}. 

\begin{figure} 
\centerline{\includegraphics[angle=0,width=\columnwidth]{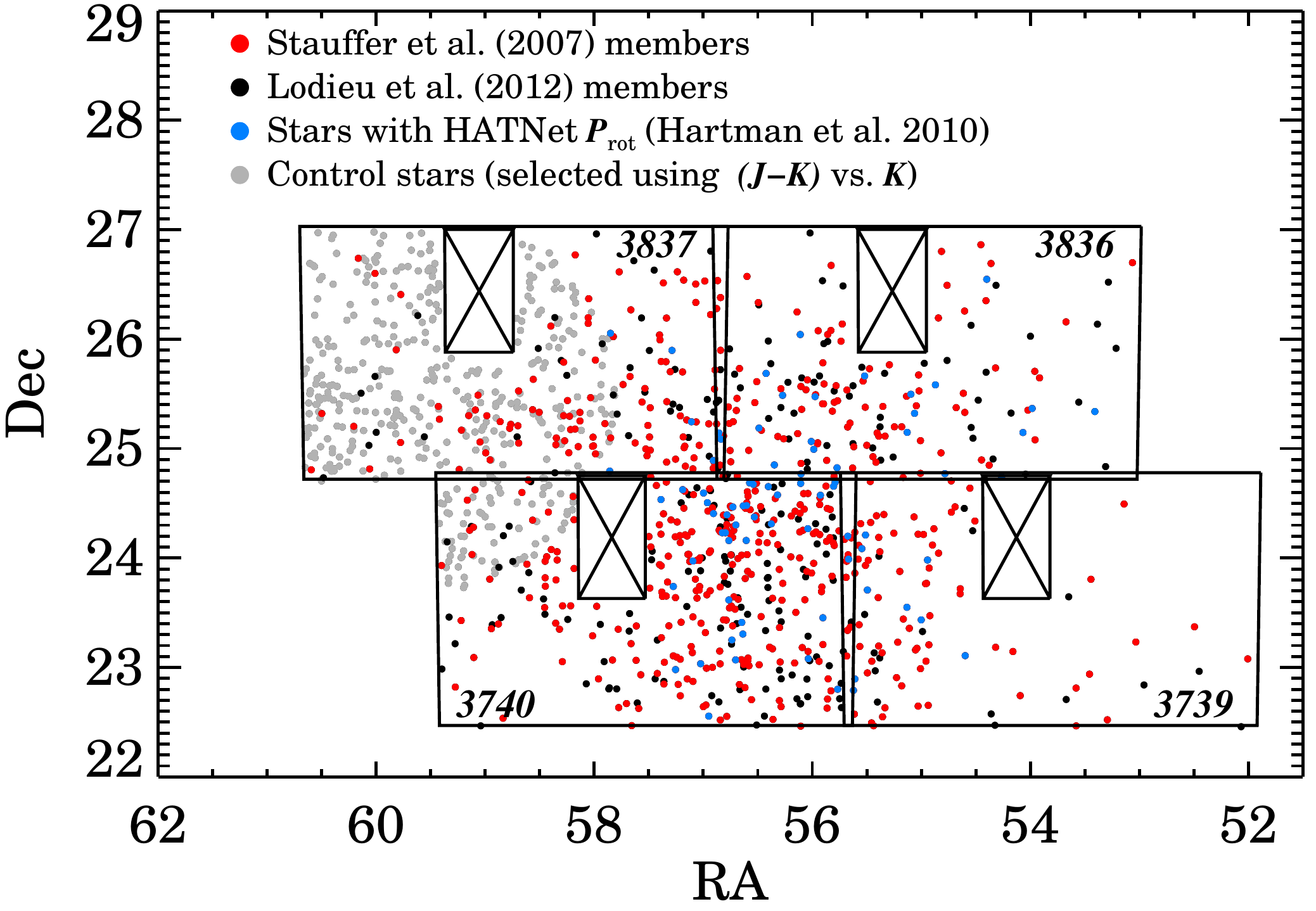}}
\caption{Distribution of probable and possible Pleiades members with 2011--2012 PTF light curves. Pleiades members compiled by \citet{Stauffer2007} are shown in red, with additional candidates compiled by \citet{Lodieu2012} in black. Stars with \Prot\ measured by \citet{Hartman2010} are highlighted in blue. A control sample of field stars that are offset from the cluster core and have PTF light curves and NIR colors and magnitudes consistent with cluster membership, is in gray. Individual PTF fields selected for POCS monitoring are overlaid in black; the location of the dead CCD in each PTF field is crossed out.} 
\label{fig:fields}
\end{figure}

$R$-band observations of the four PTF fields that contain the most candidate Pleiades members were scheduled in the {fall} and winter of 2011--2012 {(see Figure}~\ref{fig:fields}{)}. Observations began on 2011 Sep 6 and {ended} on 2012 Mar 4; the number of visits each field received is {given} in Table~\ref{tab:fields}. The $\gapprox$300 $R$-band visits to each field exceed by nearly an order of magnitude the number of visits that neighboring PTF fields received in the standard PTF survey. To provide sensitivity to long and short \Prot, each Pleiades field received a mixture of low-density ($\approx$1--2 visits per night) and high-density ($\approx$20 visits per night) monitoring; high-density coverage was obtained for various fields in 2011 Oct and Dec and 2012 Jan.

We {follow} \citet{Law2011} in assembling our photometric light curves. In brief: {we perform aperture photometry} using SExtractor \citep{Bertin96} on each IPAC-processed PTF frame \citep{Laher2014}{, whose photometric calibration is described by \citet{Ofek2012}. This generates} photometry for all objects at each epoch with approximate zero-points determined on a chip-by-chip basis using USNO-B1 \citep{Monet2003} photometry of bright stars. After {removing} observations affected by e.g., bad pixels, diffraction spikes, or cosmic rays, the positions of single-epoch detections were matched using a 2\arcsec\ radius to produce multi-epoch light curves.

{We then examine the light curves} to identify epochs where a large fraction of the objects on each CCD had anomalous photometry due to atmospheric effects (most often due to clouds) or elevated background (e.g., moonlight). After removing the <2\% of epochs that were typically flagged, as well as discounting objects exhibiting evidence for intrinsic astrophysical variability, the zero-point solution for each epoch was re-optimized to minimize the overall long-term RMS variations for the ensemble of stars. The light curves were then detrended using {five iterations of} the SysRem algorithm \citep[][]{Tamuz2005} to remove smaller-scale variations that affected groups of stars on sub-chip-scales, such as airmass variations across the image and thin small-scale clouds.

\begin{deluxetable}{lccc}[t!]
\tablewidth{0pt}
\tabletypesize{\scriptsize}
\tablecaption{PTF Observations of Pleiades Fields}
\tablehead{
\colhead{Field} & \colhead{Field Center} & \colhead{Candidates }  & \colhead{Visits} \\
\colhead{Number} & \colhead{(J2000)} & \colhead{with PTF LCs}  & \colhead{} 
}
\startdata
3739 & 03:35:15 $+$23:37:30 & 99 & 327 \\
3740 & 03:50:06 $+$23:37:30 & 389 & 287 \\
3836 & 03:39:47 $+$38:19:00 & 174 & 310 \\
3837 & 03:54:57 $+$37:19:00 & 156 & 303 
\enddata
\label{tab:fields}
\end{deluxetable}

\subsection{Measuring Rotation Periods}
We search for signatures of periodic variability in the PTF {data for 818} candidate Pleiads with median $R_{PTF}$ between 12.9 and 19 mag. After removing data points flagged as potentially spurious (e.g., because many sources at a given epoch significantly deviate from their means), we {compute} Lomb-Scargle periodograms {\citep{Scargle1982}} for each light curve using an updated version of the iterative process developed by \citet{Agueros2011}, extended by \citet{Xiao2012}, and illustrated in Figures~\ref{fig:example_periodogram1} and \ref{fig:example_periodogram2}.  

In each iteration, {the} Lomb-Scargle periodogram {is} sensitive to periods from 0.1 to 30 days {and is} used to identify the period with maximum power. The light curve {is} then phased to this period, pre-whitened by subtracting the median value of all points within 0.035 in phase, and all points that {are} 4$\sigma$ outliers to the pre-whitened light curve {are} then excluded from the next iteration. {We adopt the} period with the maximum power in the periodogram computed after three iterations of this process as the most likely \Prot.

\subsection{Establishing Criteria for Reliable Period Measurements}
\subsubsection{Internal Check: {Pleiads} versus Control {Stars}}\label{sec:controlSample}
To evaluate the reliability and robustness of our \Prot\ measurements, we {construct} a control sample of field stars with colors and magnitudes similar to those of {\it bona fide} Pleiades members. These light curves will have the same instrumental signatures as those of our Pleiades targets, but should exhibit significantly lower levels of intrinsic astrophysical variability due to their older ages and lower levels of magnetic activity \citep[as expected based on the age-activity relation; e.g.,][]{hawley1999, soderblom2001, Douglas2014}.  By injecting artificial periodic signals into these quieter light curves, we {test} our ability to accurately recover \Prot\ from light curves that reflect the exact cadence and noise properties of our Pleiades targets' light curves. 

We select stars from the 4$^{\rm th}$ U.S.~Naval Observatory CCD Astrograph Catalog \citep[UCAC;][]{zacharias2012} database within 2 degrees of RA=4 HR, DEC=$+$25.5$^\circ$ (J2000), a region on the edge of the PTF Pleiades fields. From these 19,500 stars we then {choose} those with colors and magnitudes similar to those of the Pleiades cluster sequence; specifically, we {pick} stars with:
\begin{eqnarray*}
K &>& (J-K) \times 4 + 6;\ {\rm AND} \\
K &<& (J-K) \times 4 + 8\ {\rm OR}\ (J-K) > 0.8.
\end{eqnarray*}

Two thousand twenty four stars satisfy these constraints. Of these, 427 lie within the PTF Pleiades footprint and have light curves with a median $R_{PTF}$ between 12.9 and 19 mag. We include these stars in the $K$ versus $(J-K)$ CMD shown in Figure~\ref{fig:JKvsK_CMD}, and summarize the properties of the sample in Table~\ref{tab:membership}.

\begin{figure}[t!] 
\centerline{\includegraphics[angle=0,width=3.5in]{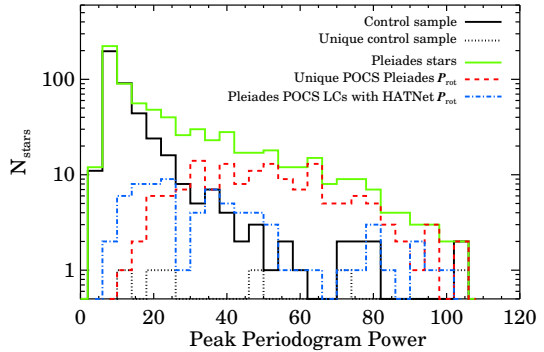}}
\caption{Distributions of the power in the primary peak in periodograms calculated from PTF light curves. Sources with strong periodogram peaks are dominated by candidate or confirmed Pleiads, which exhibit a clear excess of high-power peaks relative to the control sample of field stars. Interestingly, stars with \Prot\ reported by \citet{Hartman2010} are typically found at somewhat lower power levels than the most strongly periodic stars in the POCS sample. Stars with prominent peaks in their POCS periodogram are predominantly lower-mass stars and lie below the sensitivity of the HATNet survey, which measured \Prot\ mainly for brighter (and less photometrically variable) FGK stars. } 
\label{fig:power_comparison}
\end{figure}

To develop a catalog of robust \Prot, we {utilize} a set of objective criteria informed by previous analyses and new comparisons to empirical and synthetic period measurements. A key metric is the maximum power in the periodogram's primary peak, a quantity highlighted with an orange triangle in the middle panel of Figures~\ref{fig:example_periodogram1} and \ref{fig:example_periodogram2}. The distributions of the power in the periodogram's primary peak is shown in Figure~\ref{fig:power_comparison} for stars in the Pleiades sample and in the control sample of field stars.

The Pleiades and control samples contain similar numbers of stars whose periodograms exhibit primary peaks at low power levels (0 $<$ power $<$ 20). At powers $>$20, however, the Pleiades sample begins to exhibit a clear excess of higher-power peaks relative to what is seen in the control sample. The Pleiades sample has four times as many stars with power $\approx$30 as the control sample, and the disparity grows at higher power levels. 

The higher powers seen in the Pleiades sample are consistent with a picture in which young, magnetically active stars have higher starspot covering fractions, producing higher levels of periodic photometric variability and thus more structured periodograms than exhibited by their field-star brethren. 

In addition,
we {follow} \citet{Xiao2012} in assessing the extent to which each \Prot\ is unique and unambiguous.  We define a period to be uniquely and unambiguously measured when the periodogram contains no secondary peaks exceeding 60\% of the height of the primary peak, aside from beat periods between the primary peak and a potential 1-day alias. This threshold and the beat periods that are excluded when executing this test are identified in the middle panels of Figures~\ref{fig:example_periodogram1} and \ref{fig:example_periodogram2} with cyan and red lines, respectively. For brevity, we hereafter refer to periodogram peaks that meet this {criterion as unique}, despite the presence of many structures in the periodogram at lower levels of significance.

\begin{figure}[t!] 
\centerline{\includegraphics[angle=0,width=\columnwidth]{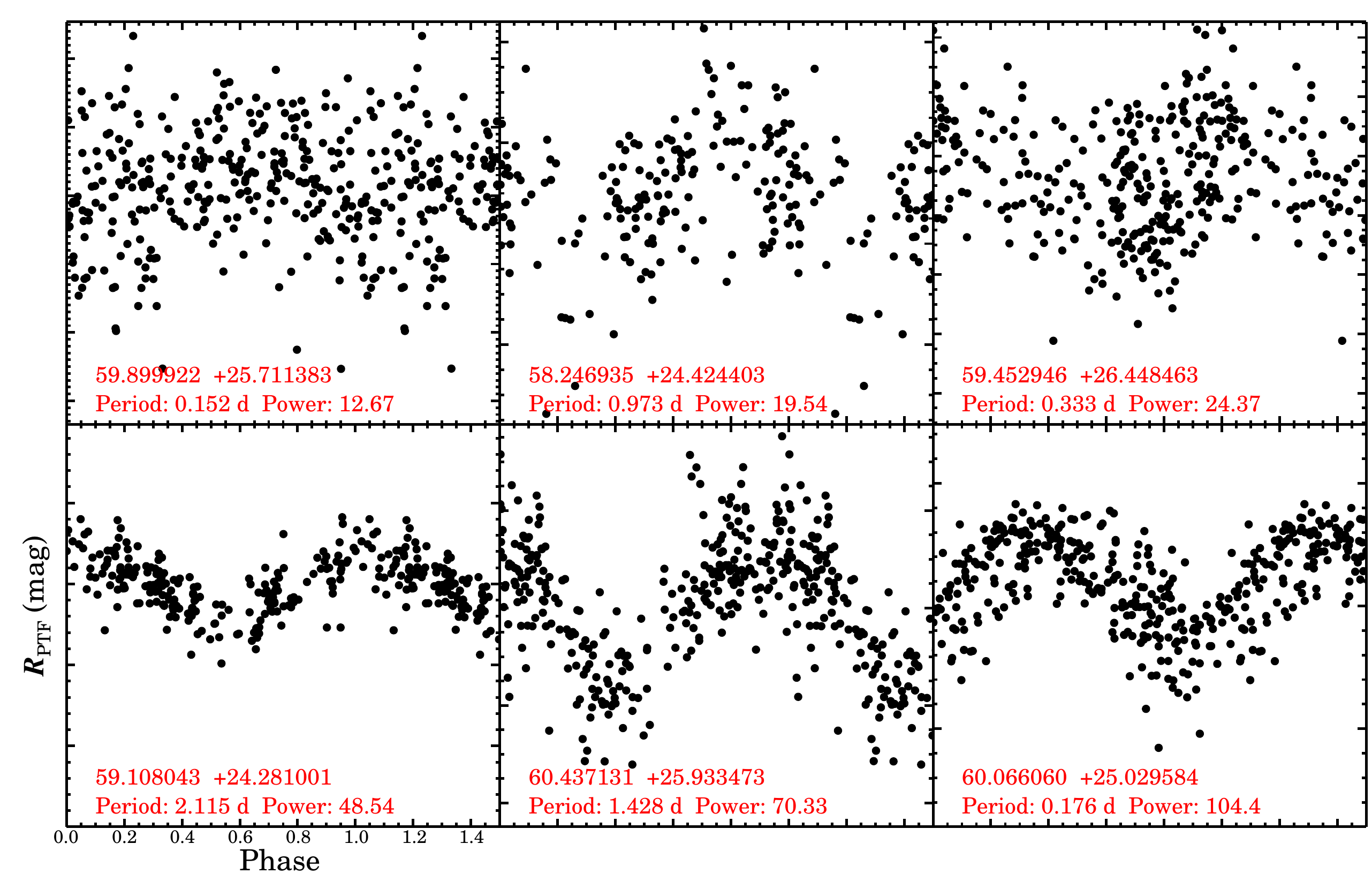}}
\caption{Phased light curves for the six stars in our control group that meet our criterion for a unique periodogram peak. The top row are the three stars with peak periodogram power $>$30; these three are clearly periodic.} 
\label{fig:Periodic_Controls}
\end{figure}

Of the {818} candidate Pleiades members with PTF light curves in our sample, 153 {produce} periodograms that satisfy this {criterion} for a unique and unambiguous \Prot\ measurement. As Figure~\ref{fig:power_comparison} shows, unique periodogram peaks are among the strongest on an absolute scale, and only six unique peaks are {found} for the control sample of older field stars in the same area of NIR color-magnitude space as that occupied by {\it bona fide} {Pleiads}. Phase-folded light curves for these six control stars are shown in Figure~\ref{fig:Periodic_Controls}. Three have periodogram peaks with powers $\leq$25; the three with power $>$30 are clearly periodic.  

\subsubsection{External Test: Comparison with HATNet \Prot\ Measurements}
\begin{figure}[t!] 
\centerline{\includegraphics[angle=0,width=\columnwidth]{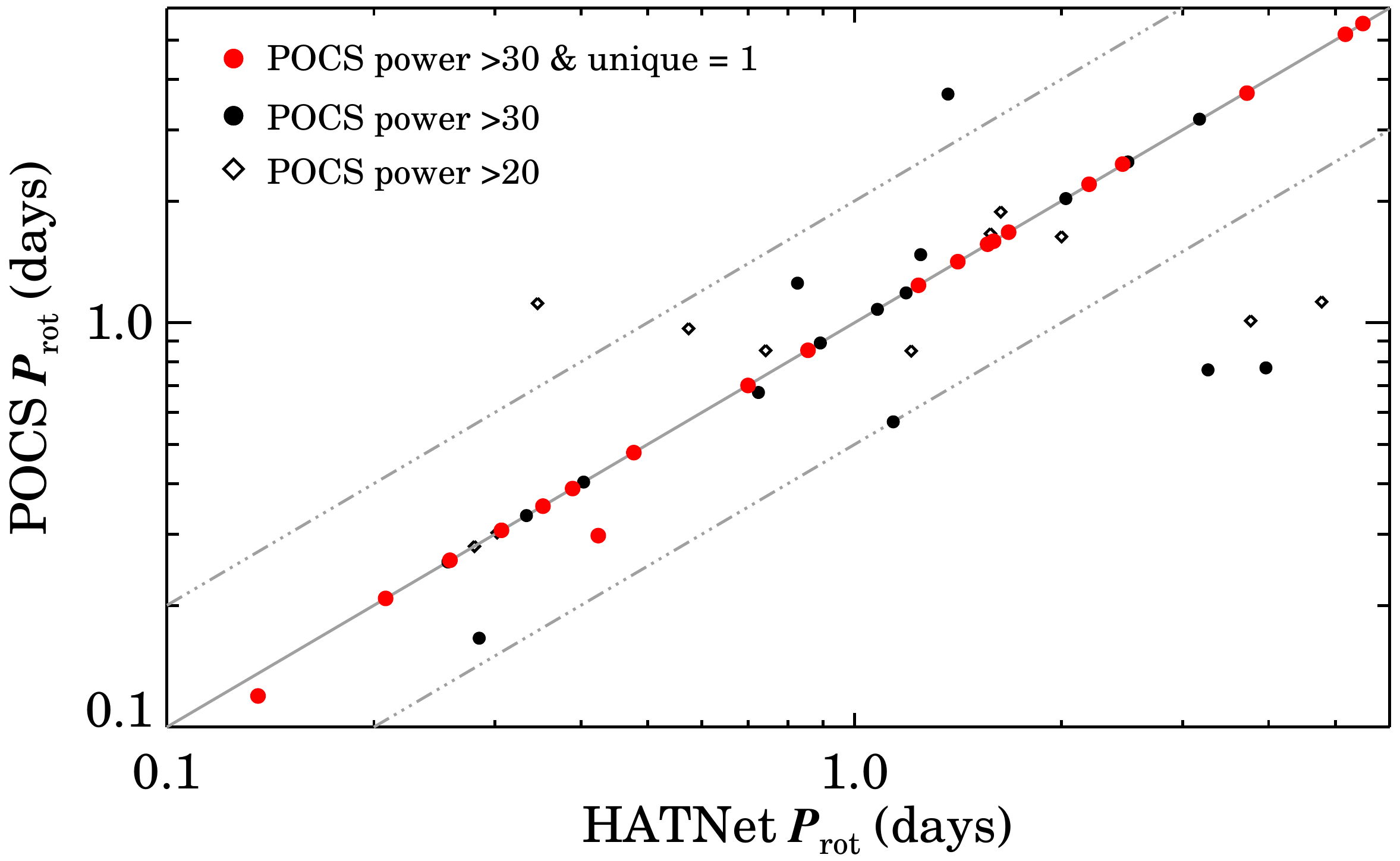}}
\caption{A comparison of the \Prot\ measured by the HATNet and PTF surveys. The solid line corresponds to a 1:1 agreement, while the dashed lines correspond to factor of 2 differences between the measured \Prot. \citet{Hartman2010} reported periods for 20 stars with POCS periodograms featuring unambiguous peaks (i.e., unique = 1), all of which also have a power $>$30. The POCS and HATNet \Prot\ are identical for 90\% (18/20) of these sources; the remaining two are rapid rotators (\Prot\ $<$ 0.5 days) where the HATNet and POCS periods differ by 0.012 and 0.12 days. The other POCS targets with HATNet periods have periodograms with strong secondary peaks (i.e., unique = 0); these stars show poorer agreement with the HATNet measurements, even when the primary peaks have absolute power levels well above the minimum power of the objects meeting the `unique' criterion.} 
\label{fig:PTFvsHAT}
\end{figure}

\begin{figure*}[t!]
\centerline{\includegraphics[angle=0,width=5.6in]{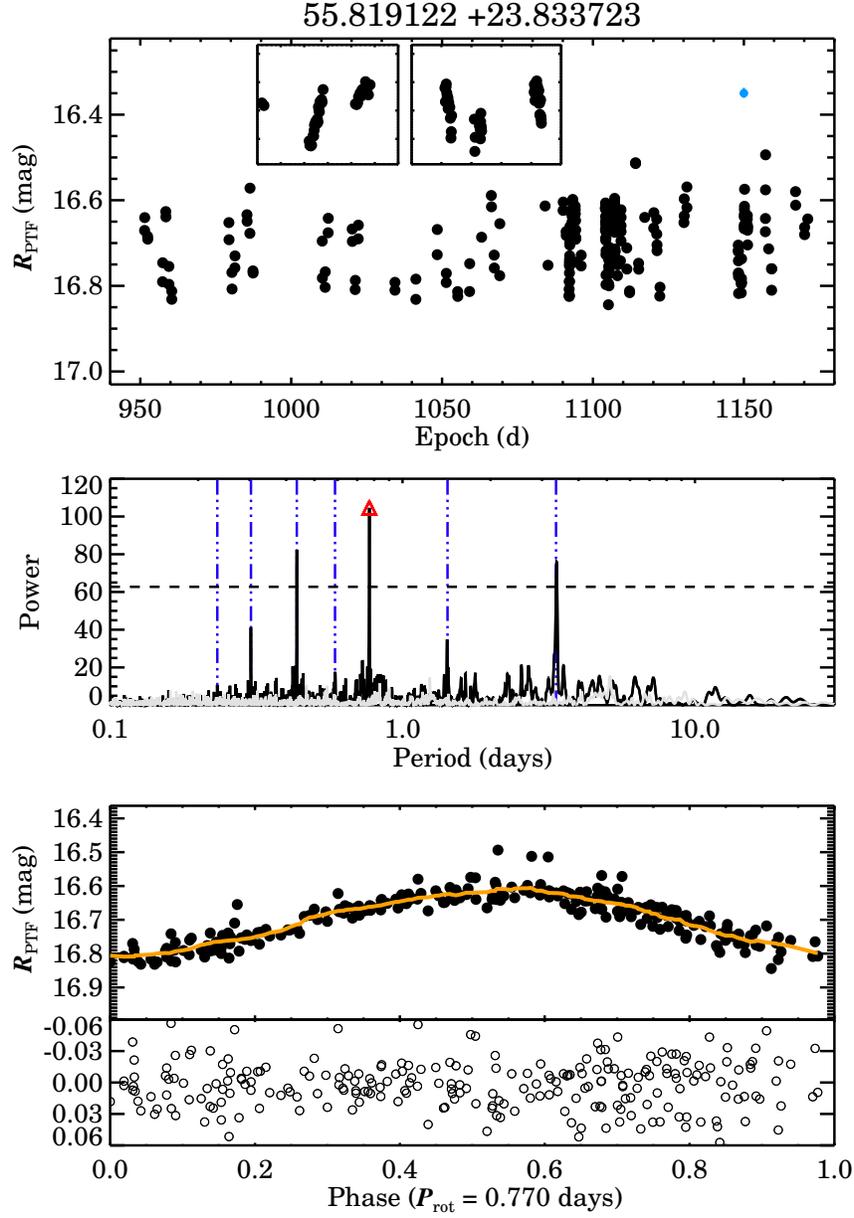}}
\caption{An illustration of our period-finding procedure. \textit{Top:} The full PTF light curve for this candidate Pleiad. The inserts correspond to times when this field was monitored at a high cadence (days 1091-1094 and 1103-1108, respectively). The light blue point shows the median photometric uncertainty on these data; here it is about the size of the data point. \textit{Middle:} The periodogram calculated from this light curve via our iterative process (black line), with the peak power, corresponding to a period of 0.770 days, highlighted with an red triangle.  Beat periods between this period and a 1-day alias are flagged with vertical (dark blue) dashed lines; the power threshold used to flag sources with ambiguous period detections (i.e., additional periods with $\geq$60\% of the primary peak's power) is shown as a horizontal dashed line. \textit{Bottom:} The light curve phase-folded to the 0.770-day period. A median-filtered version of the phase-folded light curve, shown as an orange line, is subtracted to create a pre-whitened light curve, shown in the sub-panel at the bottom. The periodogram computed from this pre-whitened light curve is shown as a gray line in the middle panel. The primary peak and beat periods are not present in the periodogram of the pre-whitened light curve, indicating that the periodic signature removed during the pre-whitening accounts for all of the significant structure in the star's light curve.\\ } \label{fig:example_periodogram1}
\end{figure*}

\begin{figure*}[t!]
\centerline{\includegraphics[angle=0, width=5.6in]{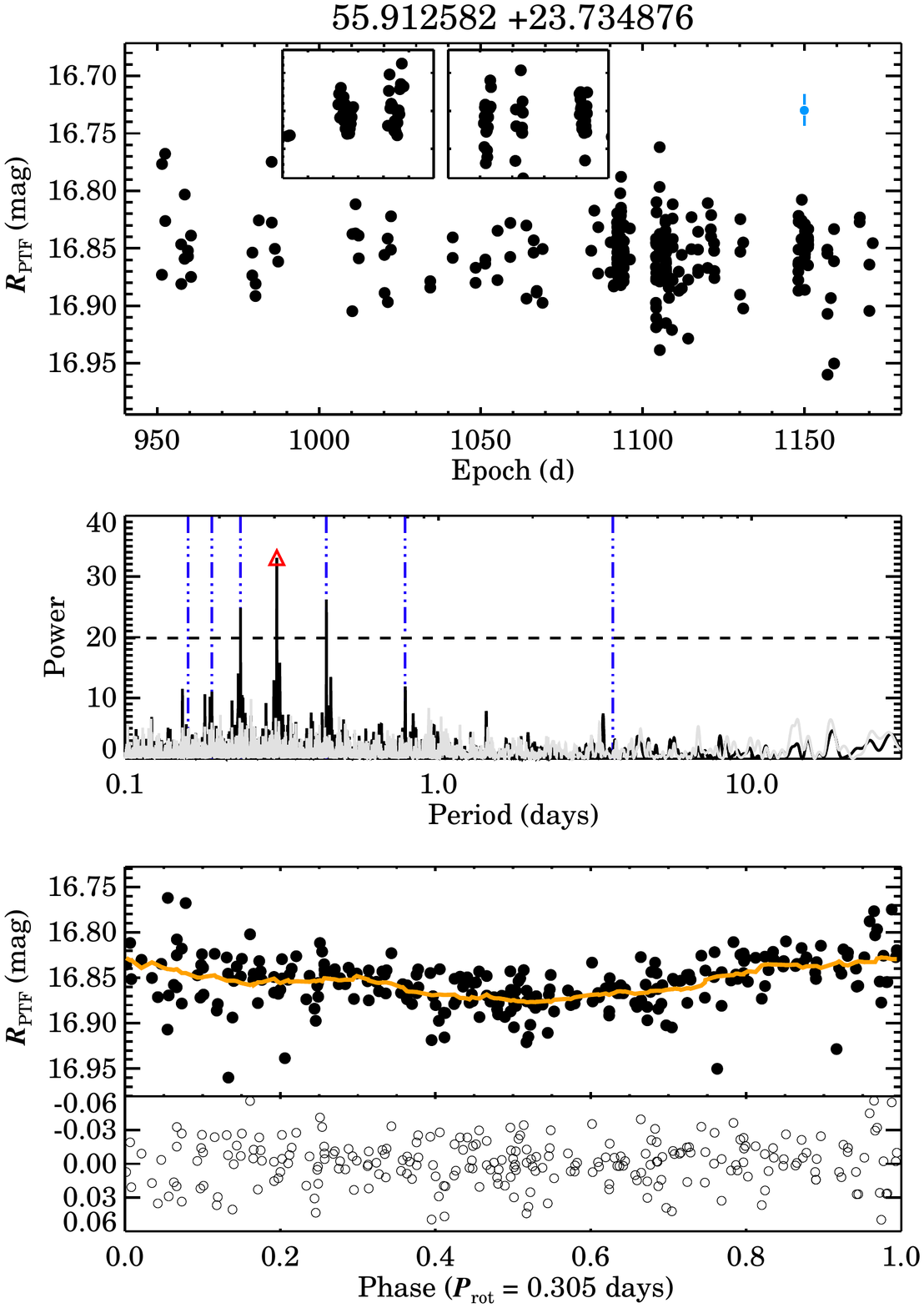}}
\caption{As in Figure~\ref{fig:example_periodogram1}, but for a star whose light curve produces a periodogram peak lying near the power cutoff defined in Section~\ref{sec:periodMeasure} for making robust \Prot\ measurements.}
\label{fig:example_periodogram2}
\end{figure*}

We {examine} the agreement between the \Prot\ we measure using POCS data and {the independent} \Prot\ {measured for Pleiads} by \citet{Hartman2010} using photometry from the HATNet planet-search program. Figure~\ref{fig:PTFvsHAT} compares the \Prot\ for 54 stars for which the POCS periodogram includes a peak with power $>$20. Twenty of these stars have a POCS periodogram with an unambiguous peak (unique flag = 1). All of those peaks have power $\geq$30, and the resulting POCS period measurement is {identical to the HATNet period} ($\Delta$\Prot $<$ 1\%) for all but two sources, representing a 90\% recovery rate. 

The stars with differing \Prot\ are fast rotators with strong periodogram peaks. Assuming the HATNet periods measured for these stars are correct, the 10\% relative disagreement measured for one star, which has a POCS power of 75, corresponds to a small absolute difference in \Prot\ (0.119 versus 0.135 days). The second star appears to lie on the locus of beat periods between true periods and a 1-day sampling cadence shown in Figure~\ref{fig:input_output}, a difficult-to-eradicate failure state for ground-based monitoring programs, as demonstrated by the Monte Carlo simulations discussed below. 

Examining the POCS \Prot\ for an additional 20 sources with power $>$30 but ambiguous period measurements (unique = 0), we {find} much poorer agreement with the HATNet results. The POCS and HATNet periods agree to better than 3\% for only 60\% (12/20) of these sources. Relaxing the criteria further, periods derived from ambiguous periodograms with peaks with power between 20 and 30 agree to within 3\% with the HATNet measurement in only $\approx$15\% (2/14) cases.

We also {examine} the overlap and agreement between our \Prot\ values and those reported by \citet{Scholz2004} for nine low-mass Pleiads. POCS {light curves} were recorded for eight of these sources, but all but one produced periodograms featuring only weak {peaks} (power $<$20). The exception {is} BPL 102, one of the most massive objects in the \citet{Scholz2004} sample. {The POCS periodogram features a moderate peak (power $\approx$}38) at 0.81 days, {consistent} with the period measured by \citet{Scholz2004}, but additional peaks {are} detected with a significant fraction of this power {(i.e., unique = 0)}, preventing the unambiguous identification of this period from the POCS {data}. 

These tests demonstrate the importance of evaluating both the absolute and relative power of a given periodogram peak in establishing its reliability.

\subsubsection{External Test: Monte Carlo Validation of Pipeline Results}
We {use} a Monte Carlo approach to verify that our pipeline accurately recovers variability signals injected into the PTF light curves of our control sample of stars. As indicated by their modest powers in Figure~\ref{fig:power_comparison}, stars in our control sample typically exhibit low levels of photometric variability compared to our sample of candidate and confirmed {Pleiads}. To ensure that any intrinsic astrophysical variability in these light curves will be dominated by the artificial signals we inject, we remove 33 stars from the control sample whose periodograms feature a peak with power $\geq$30. The light curves of the remaining 394 stars still surely contain meaningful astrophysical variability. But any degradation of the artificial signals we introduce into these light curves caused by the stars' intrinsic variability will only cause us to underestimate the performance of our pipeline, resulting in overly conservative thresholds for extracting reliable \Prot. 

\begin{figure}[t!] 
\centerline{\includegraphics[angle=270,width=\columnwidth]{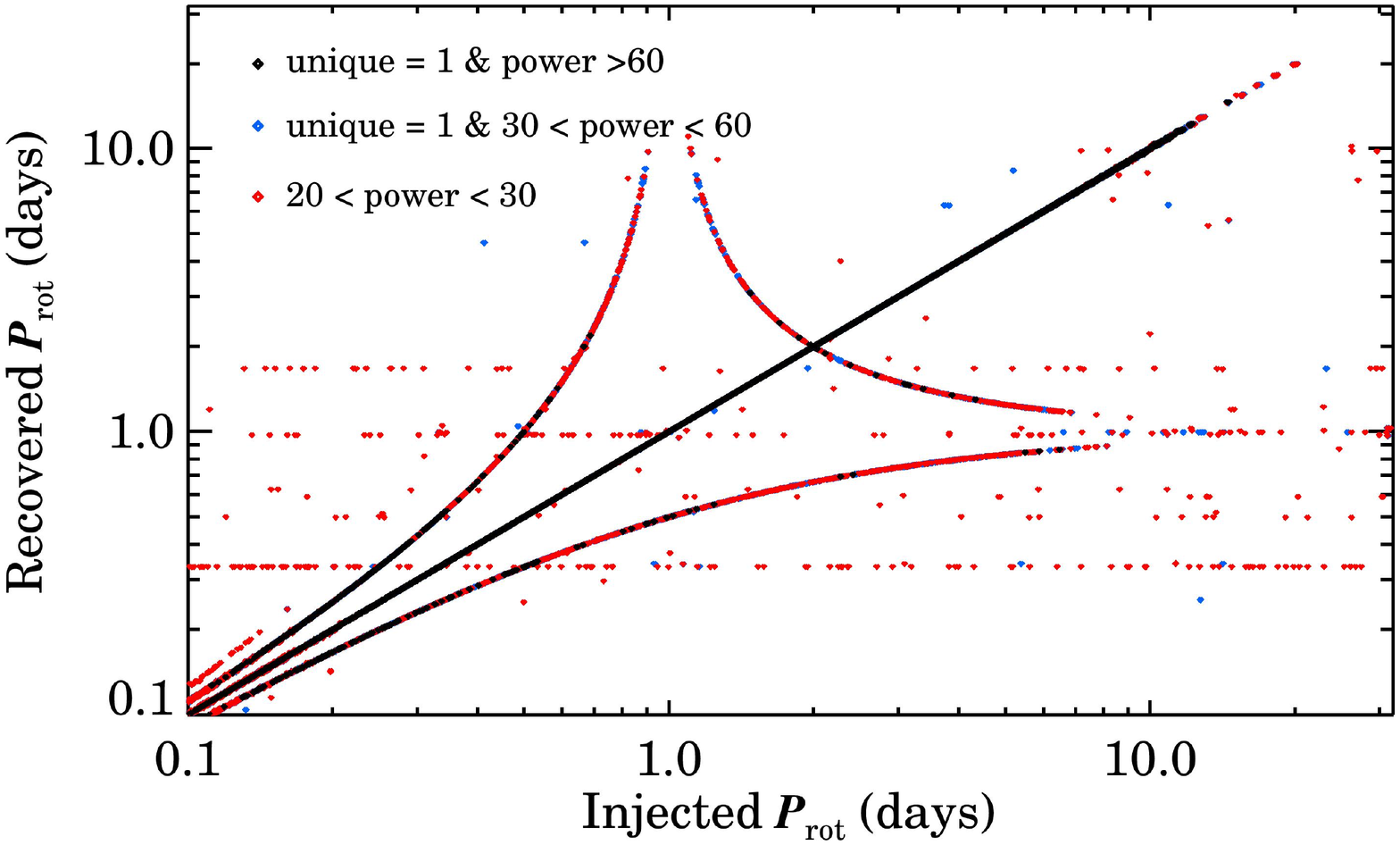}}
\caption{A comparison of \Prot\ injected into, and recovered from, PTF light curves of a control sample of 394 field stars with colors and magnitudes similar to those of {Pleiads} and lacking initial periodogram peaks with power $>$30. Spurious measurements arise primarily from beat periods between the true \Prot\ and a 1-day sampling frequency (i.e., the curved lines tending to the limit of \Prot\ $=1$ day in x and in y).} 
\label{fig:input_output}
\end{figure}

We first {create} densely sampled sinusoids with \Prot\ selected randomly from a uniform distribution in log space of $-1$~$<$~log~\Prot~$<$~1.5, corresponding to minimum and maximum periods of 0.1 and 31.6 days. The amplitude of each sinusoid {is} scaled relative to the standard deviation of the target light curve, with unique sinusoids generated for each of five different amplitude ratios: amplitude/$\sigma_{light\ curve} = $0.3, 0.6, 0.9, 1.2 and 1.5. 

{We then add these sinusoids} to the target light curve after interpolating the function onto the exact epochs for each measurement. We {generate} 500 sinusoids for each of the amplitude ratios and {test} our ability to recover 2500 unique instances of periodic variability from the light curves of each of the 394 stars in our power-restricted control sample. 

Applying our algorithm to the set of 985,000 artificially variable light curves, we {measure} the dependence of the recovery rate and accuracy of our period detection on the properties of the input light curve (i.e., period and amplitude) and the output periodogram (i.e., absolute and relative height of the periodogram peak).

\begin{figure}[t!] 
\centerline{\includegraphics[angle=0,width=.96\columnwidth]{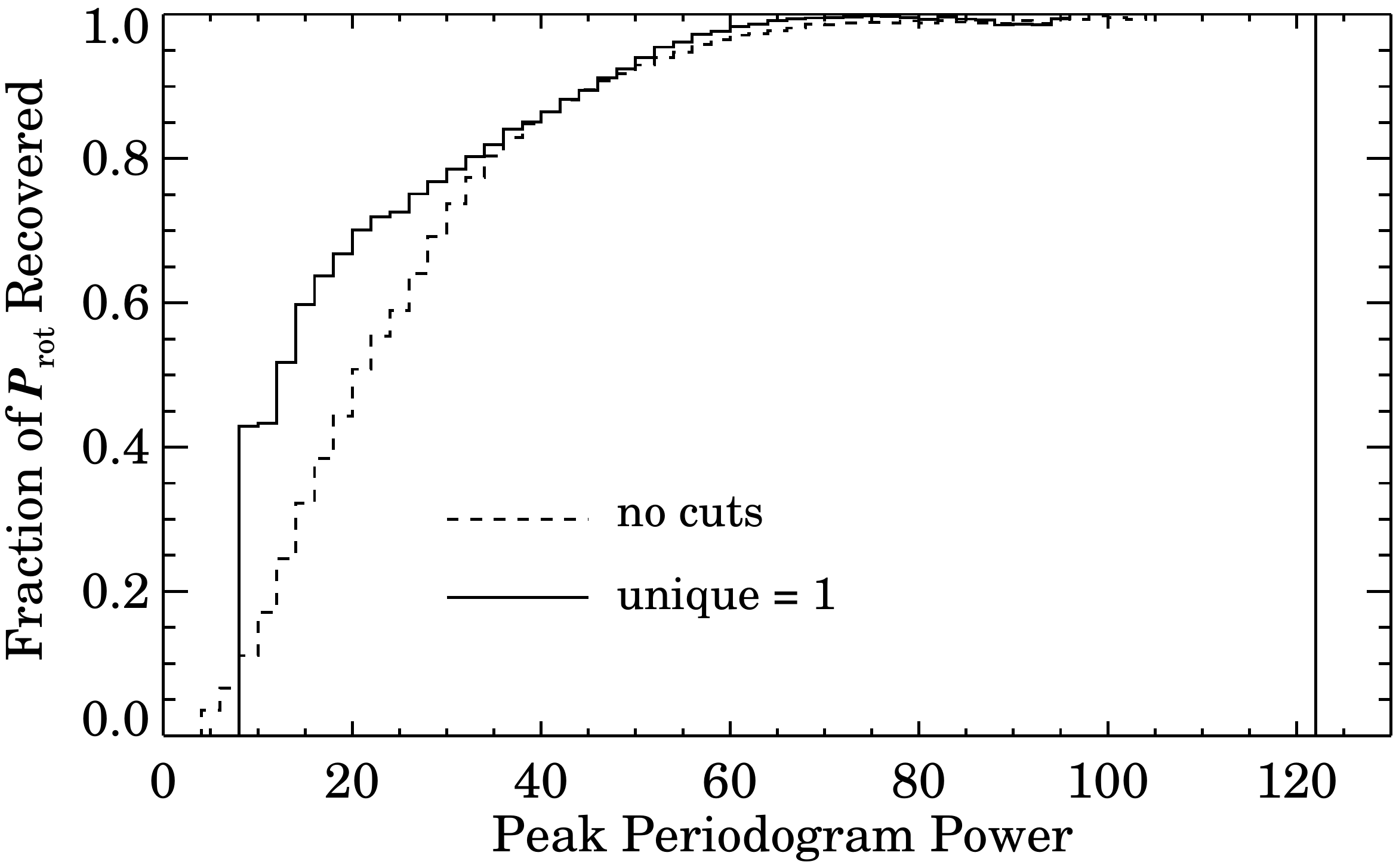}}
\caption{The fraction of successfully recovered synthetic periods as a function of the corresponding peak periodogram power. For powers $\lapprox$40, the recovery rate of the true period when the period is unambiguous (i.e., no secondary peaks with power $>$60\% of the primary peak's; solid line) is 20--25\% higher than the rate for the full sample of periodiograms (dashed line).} 
\label{fig:cum_error}
\end{figure}

In Figure~\ref{fig:input_output} we compare the simulated and recovered periods for individual light curves, and in Figure~\ref{fig:cum_error} we display the cumulative distribution function of the fraction of \Prot\ successfully recovered when applying different quality cuts to the sample. We define a successful recovery as one in which the input and recovered \Prot\ agree to within 3\%, and find that our overall success rate is 66\% for this simulation. 

Spurious measurements arise primarily from beat periods between the true \Prot\ and the typical 1-day sampling frequency of the PTF monitoring. These spurious measurements are most prevalent among, but not limited to, stars whose periodograms feature multiple strong peaks (i.e., unique = 0) or low power levels ($<$30). As Figure~\ref{fig:cum_error} shows, the recovery rate increases with the strength of the periodogram's primary peak, and restricting the sample to sources with a single strong peak (unique = 1) enables the recovery rate to exceed 50\% even for periodograms with relatively weak power levels ($\approx$20).  

\begin{deluxetable}{cc}
\tablewidth{0pt}
\tabletypesize{\scriptsize}
\tablecaption{Building a Robust Sample of Rotators}
\tablehead{
\colhead{Pleiads} & \colhead{Number} 
}
\startdata
\multicolumn{1}{l}{with PTF light curves} & 818 \\
\multicolumn{1}{l}{\hspace{0.4cm} ...and unique periodograms} & 154 \\
\multicolumn{1}{l}{\hspace{0.4cm} ...and power $>$30} & 132 
\enddata
\label{tab:PeriodicSample}
\end{deluxetable}

\subsection{New, Reliable Periods for Low-Mass Pleiads}
As a result of the work described above, we {adopt} the following criteria to define reliable \Prot\  measurements:

\begin{enumerate}
\item{\textbf{a unique periodogram peak.} We eliminate stars with periodograms that include, aside from expected beat periods, secondary peaks with power $>$60\% that of the primary peak. This requirement is motivated by the poor agreement between the POCS and HATNet periods for stars with otherwise strong periodogram peaks.}
\item{\textbf{a peak periodogram power $>$30.} We also eliminate stars whose periodograms feature primary peaks with power $<$30. This criterion is motivated by our pipeline's poor success rate ($<$80\%; see Figure~\ref{fig:cum_error}) at accurately recovering periods from simulated light curves producing unambiguous but weak (power $<$30) periodogram peaks. }
\end{enumerate}

Table~\ref{tab:PeriodicSample} summarizes the number of stars that pass each stage of these quality cuts. Our pipeline produced a {robust \Prot\ for 132} Pleiads. This sample spans a range of masses from 0.18 to 0.65 \Msun, and includes 20 stars with {$M \geq 0.45\ \Msun$} and periods previously measured by \citet{Hartman2010}. Our work thus provides new \Prot\ for {112} Pleiads (including 14 candidate binaries), the vast majority of which occupy a previously unexplored area of mass-period space for this benchmark open cluster. 

\begin{figure*}[t!] 
\centerline{\includegraphics[angle=0, width=5.5in]{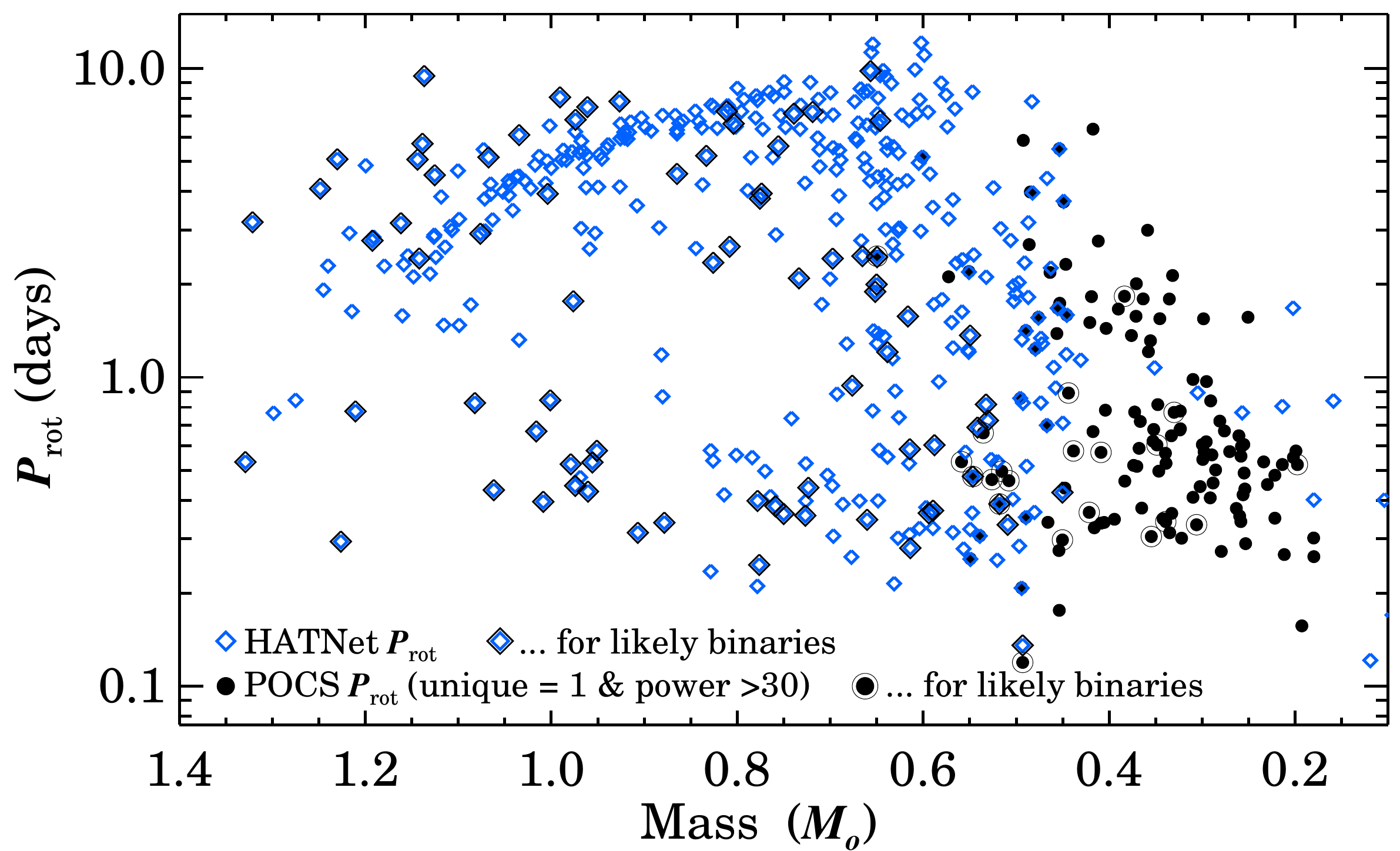}}
\caption{\Prot\ for Pleiads versus mass inferred from each star's $M_K$. PTF's sensitivity enables \Prot\ measurements for $0.18 < M < 0.45$~\Msun\ members, capturing the rapid rotation that characterizes the lowest-mass stars and complementing the extensive sample of rotation period measurements reported by \citet{Hartman2010} for higher-mass ($>$0.4~\Msun) Pleiads. The \Prot\ measured by these two programs provide complete coverage of the mass-period plane for this benchmark 125-Myr-old cluster.} 
\label{fig:MassPeriod}
\end{figure*}

We present key photometric and light curve properties for all {132} rotators in Table~\ref{tab:PeriodicSample}, with individual phased light curves shown in Figures~\ref{fig:periods_1} - \ref{fig:periods_5}.  The location of these stars in the mass-period plane is shown in Figure~\ref{fig:MassPeriod}, along with the higher-mass Pleiads with \Prot\ measured by \citet{Hartman2010}. 

The {132} \Prot\ we have obtained represent a $\approx$16\% success rate for extracting period measurements from PTF light curves of Pleiades members. Studies of older clusters return a smaller fraction of period measurements. For example, \citet{Agueros2011} used PTF to monitor Praesepe, a 600-Myr-old cluster, and obtained periods for $\approx$5\% of the cluster members that fell within the PTF fields. This is consistent with the overall decay of photometric amplitudes as a function of stellar age. 

\section{Discussion}\label{sec:discuss}
\subsection{Linking the Offsets in the $(V-K)$ versus $V$ CMD to \Prot} \label{sec:color_rotation}
Several groups {have explored anomalies} in the photometric properties of Pleiades members {\citep{Stauffer1984a,Stauffer2003,Bell2012,Kamai2014}}. These anomalies were identified as offsets between the Pleiades's cluster sequence and those measured in older open clusters (i.e., Praesepe and the Hyades) or theoretical 125-Myr isochrones. For example, \citet{Stauffer2003} found that the cluster's K dwarfs were bluer than their Praesepe analogs in the $V$ versus $(B-V)$ CMD, and redder in the $V$ versus $(V-K)$ one. In $V$ versus $(V-I)$, no offset was apparent, however, suggesting that the offsets seen in the other colors were not due to differences in the stars' $V$ magnitudes, but rather represented excesses in both $B$ and $K$. 

The photometric {anomalies seen in} Pleiades members are typically attributed to the presence of cool starspots on their stellar photospheres. \citet{Kamai2014} found evidence for this explanation in a correlation between each star's \Prot\ and its color/magnitude displacement relative to the mean cluster sequence.  Those authors interpreted this rotation-color relationship as {a signature of the increased impact of} temperature differences on the photospheres of the Pleiades's fastest-rotating, and thus most heavily spotted, low-mass members. 

\begin{figure*}[!ht] 
\centerline{\includegraphics[angle=0, width=5.5in]{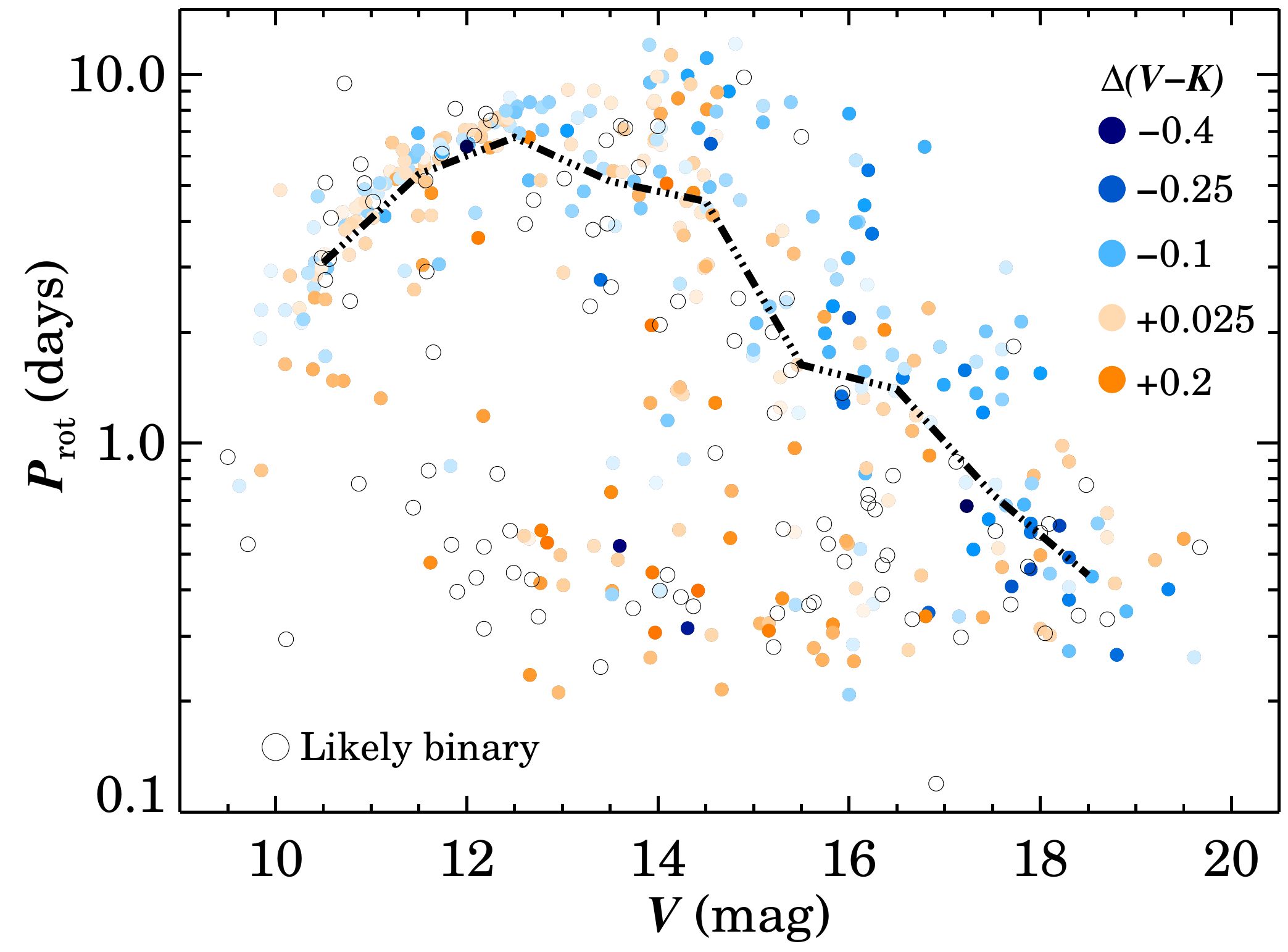}}
\caption{\Prot\ as a function of $V$ for Pleiades members, with individual points color-coded as in Figure~\ref{fig:ColorOffsets}. A color gradient is visible, particularly for sources with $V > 14$ mag. Rapidly rotating stars are systematically redder in $(V-K)$ than more slowly rotating stars with the same $V$ magnitude. To investigate the statistical significance of this effect, we overlay a dashed line to demonstrate the change in the median period as a function of $V$. We identify stars above and below this line as slow and fast rotators, respectively.} 
\label{fig:V_vs_Period}
\end{figure*}

We {utilize} our new period measurements to re-visit this potential connection between Pleiades members' \Prot\, colors and photometric amplitudes. The low-mass stars for which we measured \Prot\ are sufficiently faint and red that accurate $B$ magnitudes are difficult to acquire, as reflected by the truncation of the $(B-V)$ versus $V$ cluster sequence at $V = 17$ mag in Figure~\ref{fig:ColorOffsets}. We therefore {restrict} our analysis of these color offsets to the $V$ versus $(V-K)$ plane, where the Pleiades cluster sequence is well defined for even the faintest, lowest-mass members for which we have measured \Prot\ ($V = 19.8$, $\approx$0.18 \Msun). To {provide a simple metric for} each member's location in the CMD {relative to the cluster sequence}, we {calculate} the difference between its observed $(V-K)$ and that predicted for its $V$ magnitude by our extension of the $V$ versus $(V-K)$ cluster sequence of \citet{Kamai2014}. $\Delta (V-K)$, the distance in color space from the cluster sequence, is conveyed in Figure~\ref{fig:ColorOffsets} by the color of each point. We use this same color-coding in Figure~\ref{fig:V_vs_Period}, which shows each star's \Prot\ as a function of its $V$ magnitude. This color-coding reveals a vertical gradient in Figure~\ref{fig:V_vs_Period}, such that faster rotating stars have redder $\Delta(V-K)$ color excesses; this effect is most easily visible for 12 $ < V < $ 16, where fast and slow rotators are most widely separated. This gradient is consistent with the color-period correlation reported by \citet{Kamai2014}: slowly rotating stars have bluer $(V-K)$ colors than more rapidly rotating stars with the same $V$.

{In Figure~\ref{fig:V_vs_Period}, we also include sources that we identify as candidate binaries. As noted earlier, these are cluster members that are at least 0.375 magnitudes brighter than the cluster sequence in the $V$ versus $(V-K)$ CMD; our assumption is that an unseen secondary may be responsible for the excess $V$-band flux. Sources that are brighter than the cluster sequence for their color are also redder than the cluster sequence at their magnitude, however, so that there is likely no clear distinction in a single CMD between color anomalies due to spots and modest photometric contributions from a low-mass secondary. Indeed, our candidate binaries populate the same regions of the diagram as high $\Delta (V-K)$ sources.} 

{To make matters worse, since tidal interactions with a close companion can affect a star's angular-momentum evolution, systems with low-mass secondaries that do remain in the putatively single-star sample may contribute to the observed correlation between \Prot\ and $\Delta(V-K)$.}

{Lacking a complete census of stellar multiplicity in the Pleiades, we cannot fully disentangle the influence of binaries on the photometric and rotational signatures of cluster members.  Therefore, we first establish the statistical significance of the correlation between \Prot\ and $\Delta(V-K)$ visible in Figure~\ref{fig:V_vs_Period}, where we have removed candidate binaries with $V$ excesses greater than 0.375~mag.  We then examine how the significance of that correlation varies with the exact threshold adopted to identify candidate photometric binaries. }

{ \subsection{Statistical Significance of the Correlation Between \Prot\ and $\Delta(V-K)$} }

{To confirm the correlation between \Prot\ and $\Delta(V-K)$}, we {perform} a Kolmogorov-Smirnov (K-S) test on the $\Delta (V-K)$ distributions for rapid and slow rotators. We first {compute} the median \Prot\ for bins of $V = 1$ mag.  Using the resulting median $V$ versus \Prot\ relation, shown as a dashed line in Figure~\ref{fig:V_vs_Period}, we {divide} the sample into slow and rapid rotators {by determining if each star's \Prot\ value is larger or smaller, respectively, than the median \Prot\ for that star's magnitude bin.}

\begin{figure}[t!] 
\centerline{\includegraphics[angle=0, width=\columnwidth]{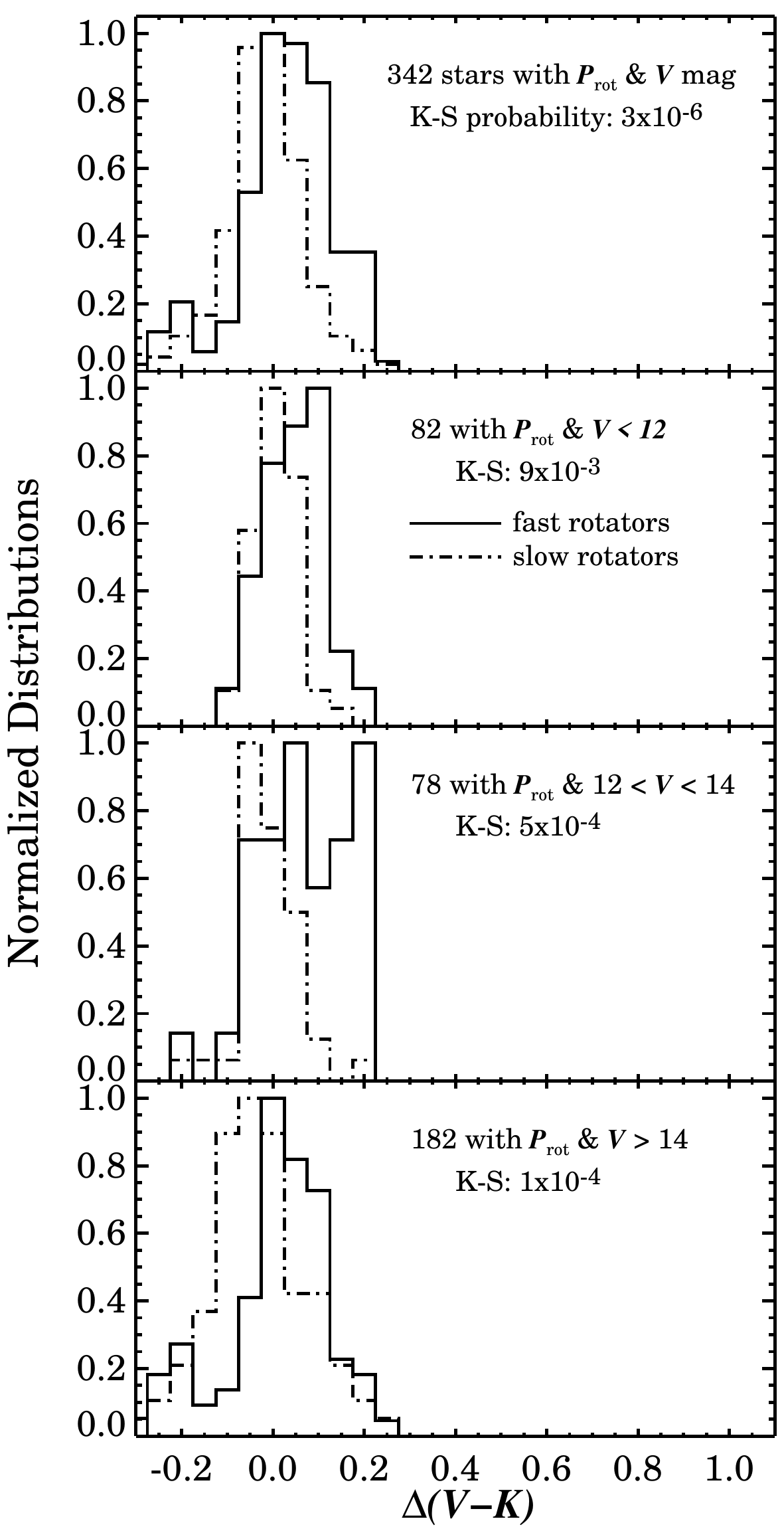}}
\caption{$(V-K)$ offsets for samples of rapid (solid histograms) and slow (dot-dashed histograms) rotators. \textit{Top:} Offsets for all Pleiads with a measured \Prot\ and $(V-K)$. \textit{Second panel:} For {82} bright ($V < 12$) Pleiads with a measured \Prot\ and $(V-K)$. \textit{Third panel:} For {78} Pleiads of intermediate brightness ($12 < V < 14$). \textit{Bottom:} For {182} faint ($V > 14$) Pleiads. In each panel, the distribution of $(V-K)$ colors is skewed to the red for fast rotators relative to the distribution for bluer, slower rotators. K-S tests indicate this effect is highly significant, with a $<$0.1\% chance that both distributions are drawn from the same parent sample in any given panel.  } 
\label{fig:delta_vk_histograms}
\end{figure}

{Figure~\ref{fig:delta_vk_histograms} shows the} $\Delta (V-K)$ distributions for rapid and slow rotators {across the sample's full range of magnitudes}, {and} for bright ($V \leq 12$ mag), intermediate ($12 < V \leq 14$), and faint subsets ($V > 14$). K-S tests strongly reject the hypothesis that the $\Delta(V-K)$ distributions for the slow and fast rotators are selected from the same parent population: in all brightness regimes, there is a $<$0.1\% chance that this is the case.

\begin{figure}[!ht] 
\centerline{\includegraphics[angle=0, width=\columnwidth]{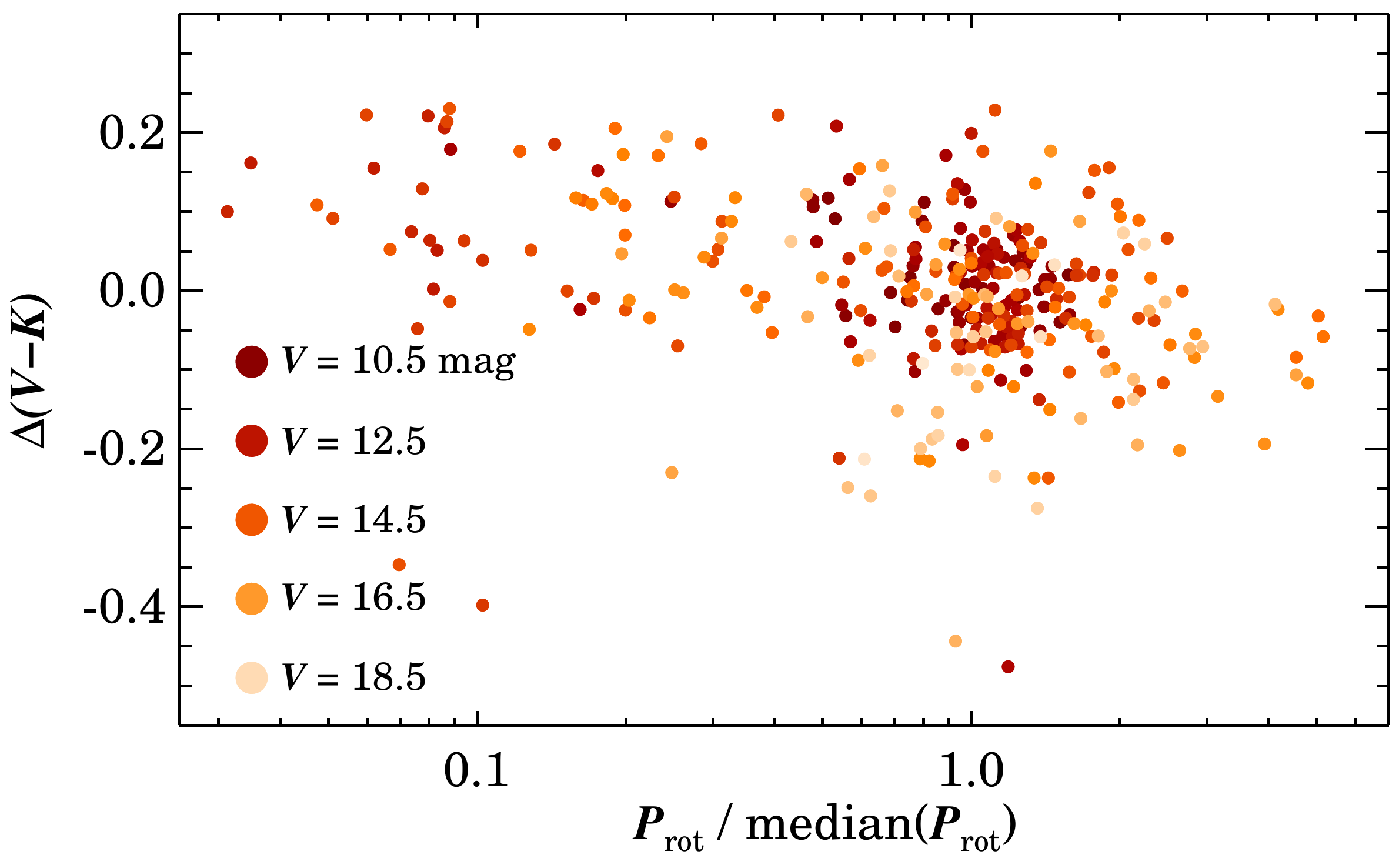}}
\caption{{$\Delta(V-K)$ offset as a function of relative rotation period for apparently single Pleiades members, with individual points color-coded according to each star's $V$ magnitude. Rotation periods are normalized by the median \Prot\ of stars within a 1 mag bin in $V$ (i.e., the points anchoring the dashed line in Figure~\ref{fig:V_vs_Period}). While there is substantial scatter, a star's \Prot\ and $\Delta(V-K)$ color offset are correlated. Rapid rotators with \Prot\ / median(\Prot) $<$ 0.3 show largely positive $\Delta(V-K)$ offsets, while slower rotators with \Prot\ / median(\Prot) $>$  1.5 exhibit mostly negative $\Delta$ (V-K) offsets. }  } 
\label{fig:VK_vs_PeriodRatio}
\end{figure}

{This correlation between rotation rate and color offset was detected by \citet{Kamai2014}, but at different significance levels for different mass regimes. Using a Spearman $\rho$ rank correlation test, these authors identified this signature for the K and M stars in their sample at a slightly higher level of statistical significance. }  
This likely reflects {the significant structure that is present in the relationship between \Prot\ and color over any significant magnitude range}. 

In the high-mass regime, {for example}, stars follow a relation between \Prot\ and mass/color/magnitude (i.e., bluer/higher-mass stars rotate more rapidly relative to redder/lower-mass stars) that directly counteracts the behavior of the rotation offset at a single mass/magnitude (i.e., rapidly rotating stars are redder than more slowly rotating counterparts at the same mass/magnitude). The opposing directions of these two effects serve to mute the overall impact of the correlation between \Prot\ and color in the mass-period plane, making the underlying correlations more difficult to detect with {the} single rank correlation test employed by \citet{Kamai2014}. 

Our approach of searching for {differences in \Prot\ and color relative to stars with similar magnitudes}, by contrast, {appears} better {suited} to detect{ing} the {\textit{second order} correlation between \Prot\ and $\Delta(V-K)$ across the full magnitude/mass range of the Pleiades CMD. Figure~\ref{fig:VK_vs_PeriodRatio} shows the correlation between a star's \textit{relative} rotation rate and color excess most directly, by plotting $\Delta(V-K)$ as a function of the star's \Prot\ value normalized to the median \Prot\ for stars in its magnitude bin. Significant scatter remains, but the fundamental relationship between a star's \textit{relative} rotation rate and color emerges: rapid rotators have positive $\Delta(V-K)$ offsets (i.e., are redder), while slower rotators have negative $\Delta(V-K)$ offsets (i.e., are bluer).}

\begin{figure}[t!] 
\centerline{\includegraphics[angle=0, width=\columnwidth]{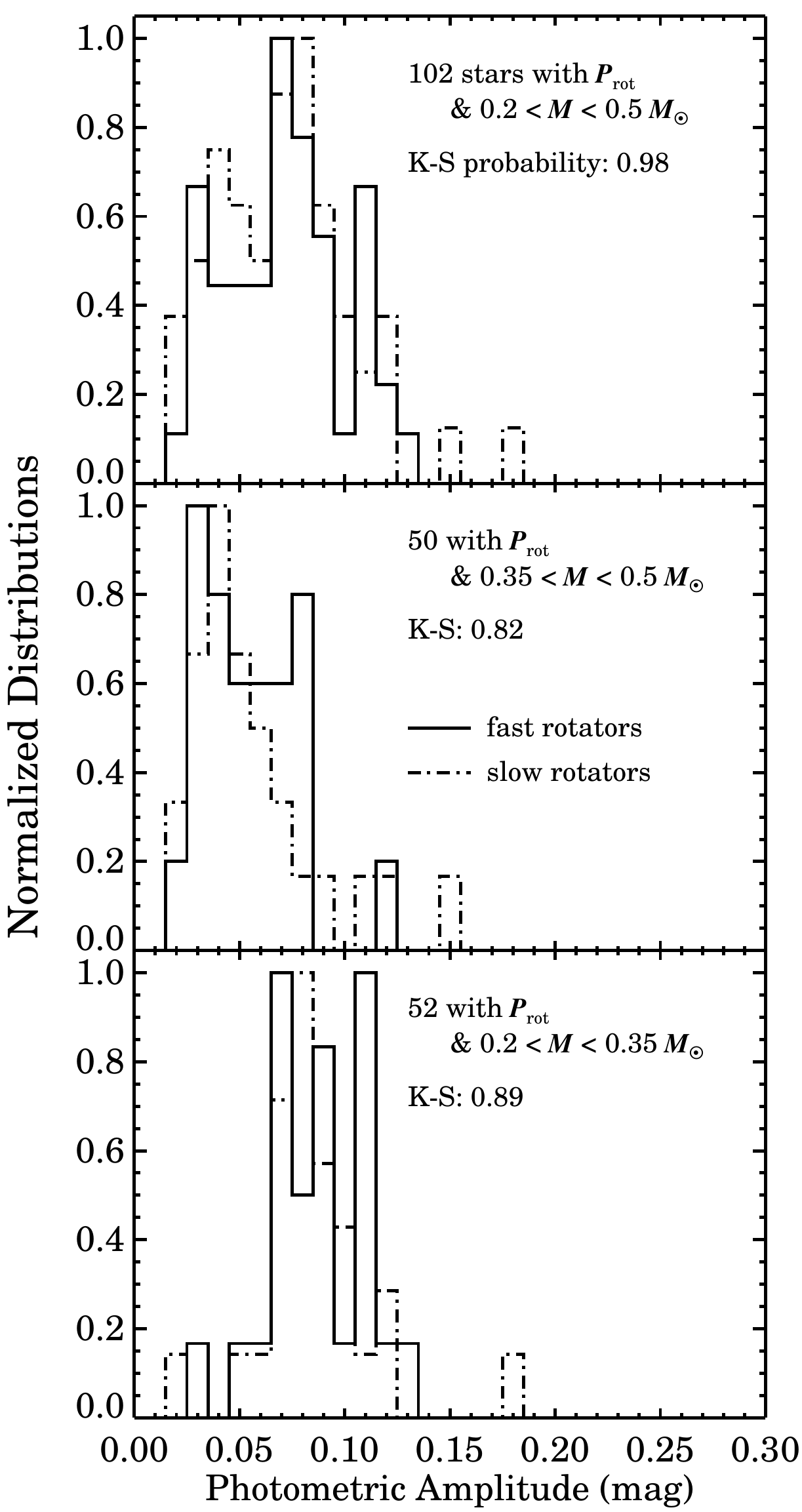}}
\caption{Photometric amplitudes measured from PTF light curves for rapid (solid histograms) and slow (dashed histograms) rotators. \textit{Top:} Amplitudes for Pleiads with masses $0.2 < M \leq 0.5$~\Msun. \textit{Second panel:} For Pleiads with $0.35 < M \leq 0.5$~\Msun. \textit{Bottom:} For Pleiads with $0.2 < M \leq 0.35$~\Msun. No significant differences are detected in the photometric amplitudes of rapidly and slowly rotating Pleiads. K-S tests indicate a {$\geq$82\% likelihood} that both distributions are drawn from the same parent sample in any given panel.  } 
\label{fig:delta_amplitude_histograms}
\end{figure}

A relationship between a star's rotation rate and the filling factor of its cool starspots could provide a natural explanation for the observed correlation between \Prot\ and photospheric colors. {As \citet{Stauffer2003} and \citet{Kamai2014} outlined previously, c}ool spots will produce redder colors, and will be more prominent on rapidly rotating stars, whose strong rotationally driven dynamos will generate large spots in regions of high magnetic flux.  As starspots are thought to be responsible for the rotationally modulated flux changes that enable \Prot\ measurements, this explanation could also imply that rapid rotators should exhibit larger photometric amplitudes than slower rotators, if the asymmetry in starspot distributions grow proportionally to the size of the spots themselves. 

We therefore searched for differences in the photometric amplitudes of the stars in our sample as a function of their \Prot. The resulting histograms are shown in Figure~\ref{fig:delta_amplitude_histograms}, divided into bins to examine the behavior across different mass regimes.  Interestingly, we find no significant difference between the photometric amplitudes exhibited by fast and slow rotators, indicating that any dependence of spot size/filling factor on rotation rate must not produce a corresponding change in the asymmetry of the longitudinal distribution of starspots. 

\begin{figure}[t!] 
\centerline{\includegraphics[angle=0, width=\columnwidth]{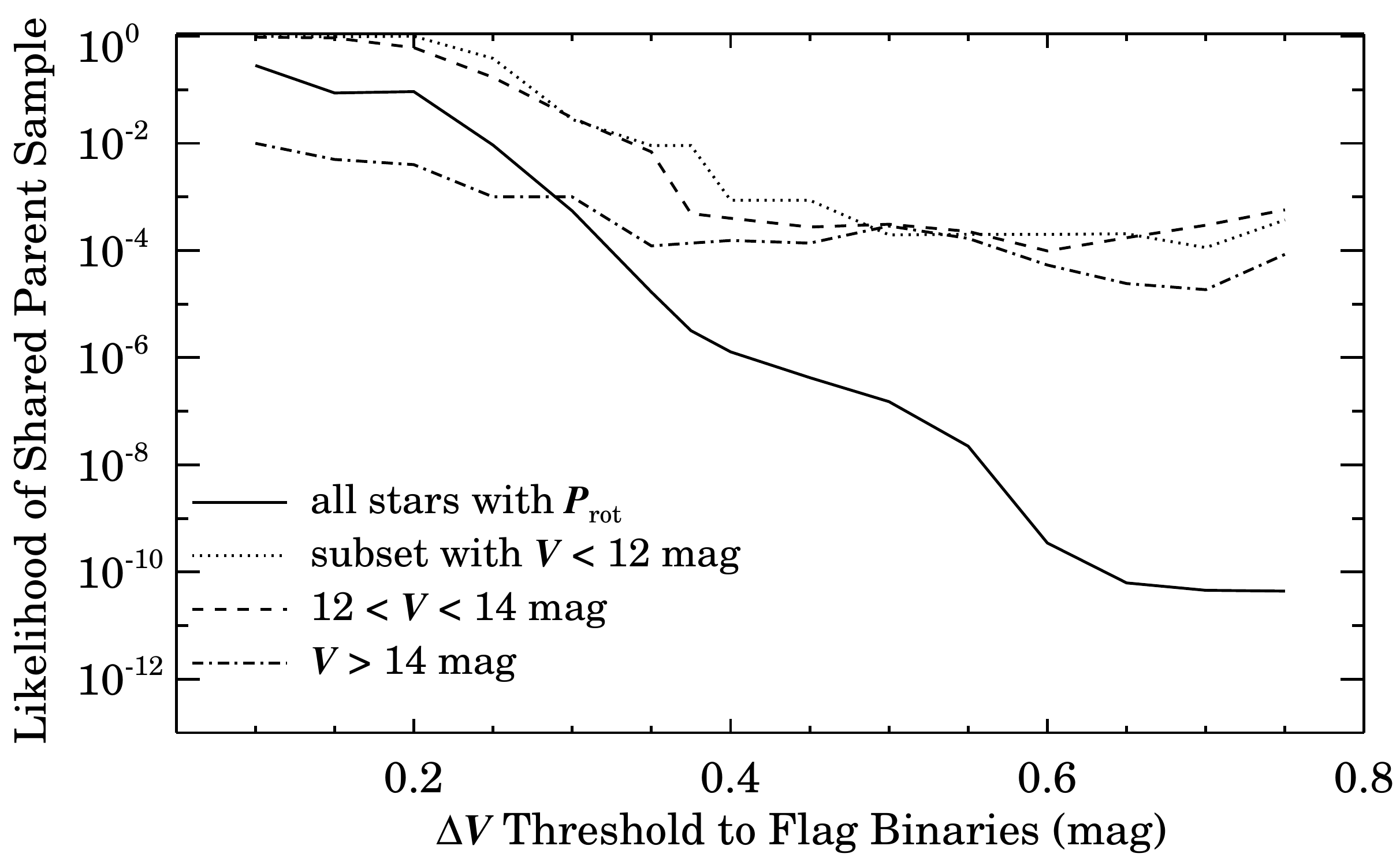}}
\caption{{Likelihood of a shared parent sample for the $\Delta$$(V-K)$ offsets of rapid and slow rotators as a function of the $V$ magnitude threshold $\Delta V$ used to remove candidate binaries from the sample.}  } 
\label{fig:signif_vs_vk}
\end{figure}

{\subsection{Sensitivity of the  \Prot, $\Delta(V-K)$ Correlation the Adopted Binary Threshold}}
{While starspots provide one explanation for the connection between a Pleiad's \Prot\ and $\Delta(V-K)$, another could be the presence and influence of an unseen secondary, which could both produce a redder $\Delta(V-K)$ offset and spin up the primary.  To test the robustness of this observational correlation against various photometric thresholds for flagging candidate binaries, we re-computed the K-S tests shown in Figure~\ref{fig:delta_vk_histograms} after using thresholds as low as 0.1 mag and as high as 0.75 mag, to remove candidate binaries from the sample. We show the resulting likelihoods in Figure~\ref{fig:signif_vs_vk} as a function of the adopted binary threshold; separate lines show the likelihoods for the full sample, and subsets of the sample drawn from narrower magnitude ranges.} 

{Adopting a stricter threshold by rejecting candidate binaries lying closer to the primary cluster sequence increases the likelihood that the $\Delta(V-K)$ distributions for the remaining fast and slow rotators are drawn from the same parent population. Increasing the likelihood of a shared parent sample for the full sample to be $>$1\%, however, requires rejecting all sources 0.25 mag or brighter than the cluster sequence as candidate binaries. And no threshold is strict enough to bring the $\Delta(V-K)$ distributions for the faintest cohort into agreement. Even rejecting stars as little as 0.1 mag above the cluster sequence results in a $<$1\% likelihood of a shared parent $\Delta(V-K)$ distribution for the faint rapid and slow rotators.}

 {Relaxing the binary selection threshold, by contrast, only increases the discrepancy between the $\Delta(V-K)$ distributions of fast and slow rotators. Objects flagged as binaries are, by definition, those with the greatest separations from the cluster sequence, and relaxing the binary threshold only adds sources with large, positive $\Delta(V-K)$ values. Furthermore, as expected if unseen companions are spinning up the primaries, Figure~\ref{fig:V_vs_Period} shows that sources flagged as binaries using our default 0.375~mag threshold are overwhelmingly fast rotators. The result is that the high $\Delta(V-K)$ sources that are added by relaxing the binary threshold are nearly all incorporated into the fast rotating population, thus enhancing the underlying $\Delta(V-K)$ discrepancy.}
 
{Ultimately, changing the threshold used to flag likely binaries does not affect the underlying empirical correlation between \Prot\ and $\Delta(V-K)$ across the full population of Pleiades members.  Adopting a strict binary threshold simply relegates the fastest rotators into the cluster's binary population, for which rapid rotation is explained as the product of interactions. Conversely, relaxing the threshold incorporates increasing numbers of rapidly rotators into the cluster's putatively single-star population, for which color excess is explained as the signature of starspots. }

\subsection{Evolution of the Mass-Period Relation}
As the lowest-mass members of the Pleiades have only recently arrived on the zero-age main sequence, their \Prot\ measurements provide a new opportunity to test whether models can correctly predict the rotational evolution of these stars. {We begin} by selecting 75 Pleiads with HATNet or POCS \Prot\ measurements {(19 from the former survey, 56 from the latter)}, no evidence of potential binarity {based on their position in the cluster CMD}, and masses $0.3 \leq M < 0.5$~\Msun.

We {find} median, 10$^{\rm th}$, and 90$^{\rm th}$ percentile \Prot\ values of {1.21}, 0.34, and 3.70 days, respectively. Following \citet{Agueros2011}, we then {use} the formalism developed by \citet{barneskim2010} and \citet{barnes2010} to find the corresponding zero-age-main-sequence periods ($P_o$) for each of these representative 125-Myr-old stars, to which we {assign} a mass of $0.4$ \Msun. This $P_o$ {is} fed back into the models to predict the \Prot\ of these representative stars at ages ranging from 30 Myr to 10 Gyr.

\begin{figure}[t!] 
\centerline{\includegraphics[angle=0, width=\columnwidth]{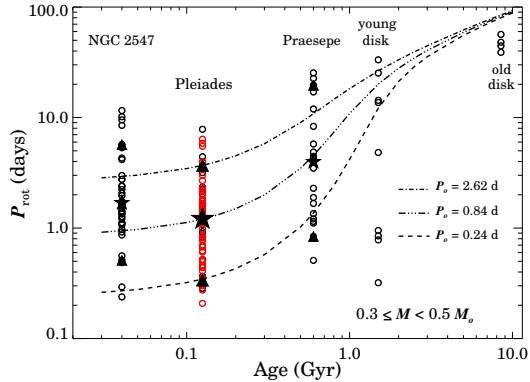}}
\caption{Predicted evolutionary tracks for $0.4~ \Msun$ Pleiads generated using the \citet{barnes2010} models compared to the observed distributions of \Prot\ for $0.3 \leq M < 0.5$~\Msun\ populations of different ages. The tracks are anchored to the median (star), 10$^{\rm th}$, and 90$^{\rm th}$ percentile (triangles) Pleiads, which are obtained from the distribution of \Prot\ for the 75 cluster members in this mass range with no evidence of potential binarity and HATNet or POCS \Prot\ measurements (black and red circles, respectively). The models reproduce the spin-down seen for the median rotator between the age of the Pleiades and that of Praesepe. However, the range of periods observed at 600 Myr is larger than that expected based on the tracks anchored by the Pleiades 10$^{\rm th}$ and 90$^{\rm th}$ percentile rotators. A comparison to the younger NGC 2547 stars shows that these stars are still spinning up before reaching 100 Myr. \label{fig:rotational_evolution} }
\end{figure}

The resulting evolutionary tracks are plotted in Figure~\ref{fig:rotational_evolution}, along with {periods} for $0.3 \leq M < 0.5$~\Msun\ stars from NGC 2547 \citep[$\approx$40 Myr; data from][]{irwin2008b}, Praesepe \citep[$\approx$600 Myr;][]{Agueros2011}, and young and old disk stars \citep[1.5 and 8.5 Gyr;][]{kiraga2007}. The masses for NGC 2547 stars were obtained by \citet{irwin2008b} using model isochrones from \citet{baraffe1998} and \citet{chabrier2000}. For Praesepe, \citet{Agueros2011} used both the empirical \cite{delfosse2000} and the theoretical \citet{Dotter2008} absolute magnitude-mass relation to obtain masses from the stars' $M_K$. Finally, \citet{kiraga2007} estimated masses for their stars based on the \citet{delfosse2000} relation for $M_V$.

The models accurately reproduce the spin-down seen for the median $\approx$0.4~\Msun\ rotator between the age of the Pleiades and that of Praesepe. However, the predictions for the 10$^{\rm th}$ and 90$^{\rm th}$ percentile stars fare less well, as the \Prot\ distribution in Praesepe is broader than one would predict based on the Pleiades data. There are a number of possible explanations for this discrepancy. One is that the models may not account correctly for the spin-down for the fastest/slowest rotators. Conversely, our sample of Pleiades rotators may be incomplete, particularly for the always-hard-to-obtain longest \Prot. Figure~\ref{fig:MassPeriod} shows a dearth of $>$3 day periods for the 0.3--0.4 \Msun\ stars, and this may explain the apparent excess of slow rotators in Praesepe relative to predictions based on our sample of Pleiades \Prot. Similarly, the HATNet and POCs sampling may not be sufficient to detect all of the fast rotators in Pleiades, leading to an apparent excess of fast rotators in Praesepe.

The disagreement between the predictions and what is seen at ages $>$1 Gyr is a by-product of the nature of the available sample of field stars. As pointed out in \citet{Agueros2011}, the \citet{kiraga2007} sample is selected from X-ray-luminous stars and therefore potentially biased toward faster rotators. In addition, these are not true single-age populations. If the fast rotators are all younger than the slow ones, the disagreement with the evolutionary tracks is not as significant as suggested by Figure~\ref{fig:rotational_evolution}. The limitations of this comparison underlines the ongoing need for (admittedly difficult-to-obtain) \Prot\ measurements of low-mass stars in older clusters.

The evolutionary tracks are extended to 30 Myr purely for reference. Low-mass stars at these young ages are still spinning up, presumably because they are still contracting. Indeed, the transition from spinning up to spinning down should occur at about the age of the Pleiades for $\approx$0.4~\Msun\ stars \citep{reiners2012}. The distributions of \Prot\ for NGC 2547 and the Pleiades shown in Figure~\ref{fig:rotational_evolution} are fully consistent with this picture.

\section{Conclusions}\label{sec:concl}
\begin{enumerate}
\item{We present results from a PTF monitoring campaign to measure rotation periods for low-mass  ($0.18 < M < 0.65$ \Msun) Pleiades members. This campaign, carried out over the fall and winter of 2011--2012, obtained $\approx$300 epochs of $R_{PTF}$ photometry for {818} Pleiades members in the cluster's central $\approx$24 square degrees.}
\item{Applying quality cuts informed by internal and external tests, we used an automated analysis pipeline to measure rotation periods for {132} Pleiades members. The periods produced by this pipeline were validated against period measurements previously reported for a subset of our sample, and results of Monte Carlo simulations where $\approx$10$^6$ synthetic periods were injected into authentic PTF light curves. These tests demonstrate that, with simple criteria to identify strong and unambiguous periodogram peaks, this pipeline is able to accurately identify a star's rotation period with $\geq$80\% reliability, depending on the strength of the periodogram peak in question.}
\item{These \Prot\ measurements reveal the morphology of the Pleiades's mass-period plane down to masses as low as 0.18 \Msun. Three-quarters {(99/132)} of the periods measured here were obtained for stars with masses $\leq$0.45 \Msun, a $>$6x increase over the number of periods that had been measured previously for Pleiads in this mass regime. These measurements demonstrate that low-mass Pleiades members occupy a distinct space in the mass-period plane, between a locus of rapid rotators with \Prot\ $\approx$ 0.25 days, and a strongly mass-dependent upper envelope of slow rotators, with the maximum period declining from $\approx$4 days at 0.5 \Msun\ to only $\approx$0.5 days at 0.2 \Msun.}
\item{When tested against rotation periods measured in the Pleiades and Praesepe, models developed by \citet{barnes2010} to describe stellar rotational evolution can quantitatively reproduce the spin-down of a typical $0.3 \leq M < 0.5$ \Msun\ star from 125 to 600 Myr. The spin-down rates predicted by the \citet{barnes2010} models do not agree quantitatively with the periods measured for rapid and slow rotators in these clusters, however. When anchored by the 10$^{\rm th}$ and 90$^{\rm th}$ percentile \Prot\ values measured for Pleiads in this mass range, models predict a narrower range of periods than is actually observed in the $\approx$600 Myr Praesepe cluster. This model-data discrepancy points to either missing physics in the rotational models, or to lingering incompleteness and bias in the samples of measured \Prot. }
\item{We confirm that rapidly rotating stars exhibit systematically redder $(V-K)$ colors than their more slowly rotating cousins. K-S tests indicate a $<$0.1\% likelihood of a common $(V-K)$ distribution for stars with \Prot\ greater and less than the median \Prot\ for their mass; this finding holds true when the cluster is considered as a whole, and when evaluating subsets covering a more restricted range of masses.  {The statistical significance of these photometric differences can be minimized if we adopt a conservative photometric binary threshold, thereby flagging most of the rapid rotators as likely binaries. In this scenario, the underlying photometric differences are explained as a product of a strong relationship between stellar rotation rate and binary frequency, rather than a dependence of photospheric/spot properties on rotation rate for putatively single rotators. } }
\item{We identify no significant difference in the $R_{PTF}$-band photometric amplitudes of slow and rapid rotators. K-S tests indicate a {$\geq$82}\% likelihood that the observed amplitude distributions could arise even if slow and rapid rotators were randomly sampling of the same parent distribution. This null detection indicates that asymmetries in the longitudinal distributions of starspots do not scale strongly with stellar rotation period; {subtler correlations may be detectable with Kepler/K2 light curves, however, given their significantly denser sampling and higher photometric precision than the PTF light curves that we analyze here.} }
\end{enumerate}

\section*{Acknowledgments}
We are grateful to Eran Ofek for his help {scheduling and carrying out the PTF observations}, to Adrian Price-Whelan for his help extracting light curves from the PTF database, and to Aaron Dotter for providing {a} 125 Myr isochrone with colors computed in the specific filters {used} during the course of this analysis. We also thank John Stauffer, Luisa Rebull, {Kristen Larson, and the anonymous referee} for comments that improved {our analysis and presentation of our results}. 

K.R.C.~thanks Hilary Schwandt and the WWU Faculty Research Writing Seminar for sage advice that accelerated the completion of this manuscript, and acknowledges support provided by the NSF through grant AST-1449476. M.A.A.~acknowledges support provided by the NSF through grant AST-1255419.

Observations were obtained with the Samuel Oschin Telescope as part of the Palomar Transient Factory project, a scientific collaboration between the California Institute of Technology, Columbia University, Las Cumbres Observatory, the Lawrence Berkeley National Laboratory, the National Energy Research Scientific Computing Center, the University of Oxford, and the Weizmann Institute of Science. 

The Two Micron {All-Sky} Survey was a joint project of the University of Massachusetts and the Infrared Processing and Analysis Center (California Institute of Technology). The University of Massachusetts was responsible for the overall management of the project, the observing facilities and the data acquisition. The Infrared Processing and Analysis Center was responsible for data processing, data distribution and data archiving. 

This research has made use of NASA's Astrophysics Data System Bibliographic Services, the SIMBAD database, operated at CDS, Strasbourg, France, and the VizieR catalogue access tool, CDS, Strasbourg, France \citep{Ochsenbein2000}. 

This research was made possible through the use of the AAVSO Photometric All-Sky Survey (APASS), funded by the Robert Martin Ayers Sciences Fund. 

\renewcommand{\thesection}{A\arabic{section}}
\setcounter{section}{0}  
{\section*{Appendix: Interesting Variable Stars in the PTF Pleiades Fields}}
{Our light-curve analysis was largely restricted to stars identified previously as candidate Pleiades members. These are a small fraction of the objects in our target fields, however, so that many other variable stars are likely to be present in the full catalog of PTF light curves. We therefore performed a broader search for high-confidence variables within the full catalog of light curves in the PTF Pleiades fields. } 

\begin{figure}[t!] 
\centerline{\includegraphics[angle=0,width=\columnwidth]{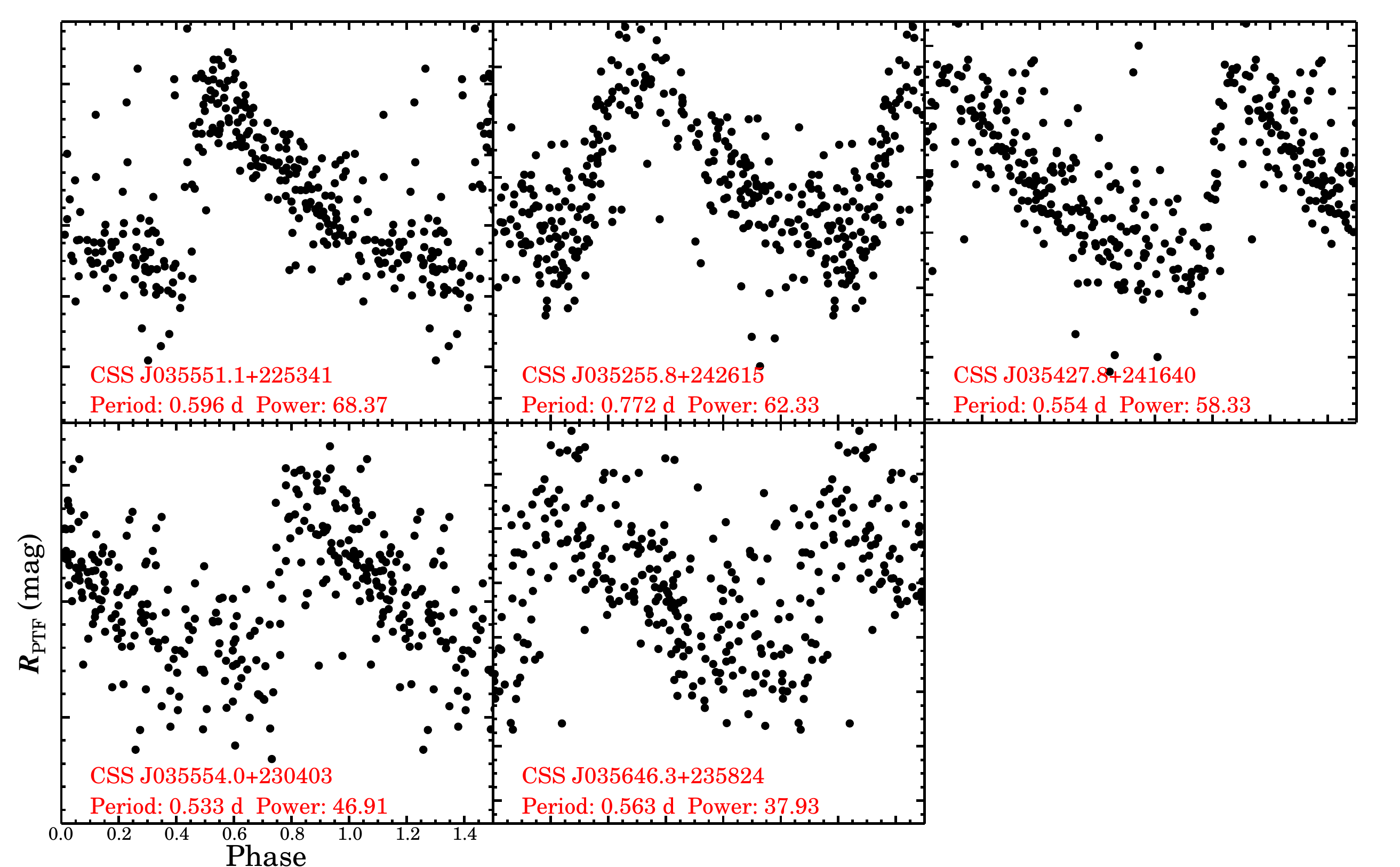}}
\caption{{Phased PTF light curves for five known RR Lyrae in our Pleiades fields.}} 
\label{fig:rr_lyrae}
\end{figure}

{Candidate variables were identified using the same period-finding algorithm described above, with the same criteria for significance and uniqueness. Among the 119 variable stars we identify in this manner are five known RR Lyrae (see Figure~\ref{fig:rr_lyrae}). All 119 stars are tabulated in Table~\ref{tab:appendix}.}


\setlength{\baselineskip}{0.6\baselineskip}

\setlength{\baselineskip}{1.667\baselineskip}

\begin{figure*}[t!] 
\centerline{\includegraphics[angle=0, width=6.5in]{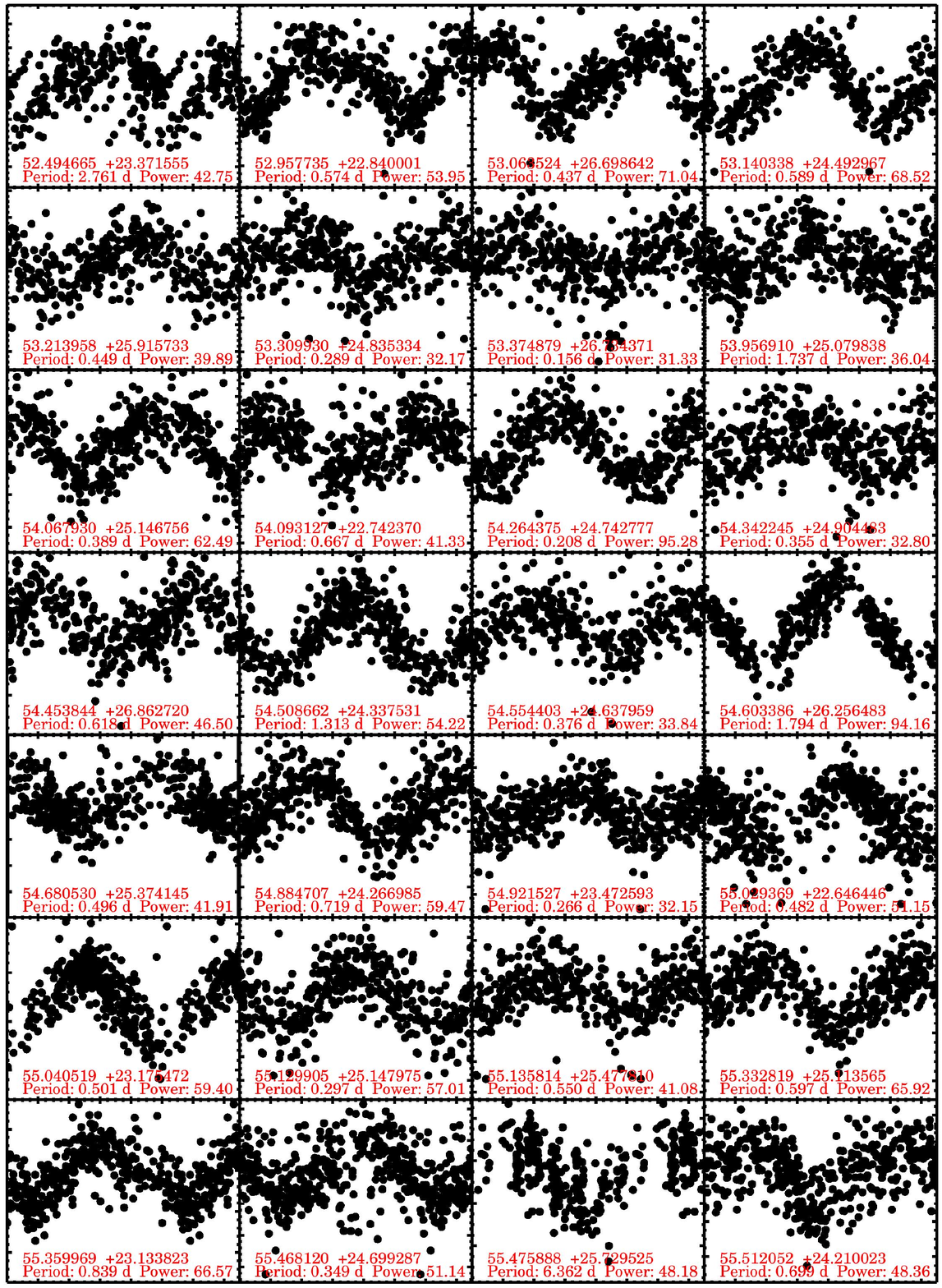}}
\caption{{Phase-folded light curves for Pleiades members with newly measured \Prot. The X axis for each panel spans 0 to 1.5 in phase; the Y axis is scaled to show 6.25 times the light curve's standard deviation, with the light curve offset vertically by 10\% to leave white space for the coordinate and \Prot\ information displayed at the bottom of each panel.  Panels are ordered according to the object's right ascension (RA), as in Table~\ref{tab:Periods}: RA increases to the right in each row, and from the top row down.} } 
\label{fig:periods_1}
\end{figure*}

\begin{figure*}[t!] 
\centerline{\includegraphics[angle=0, width=6.5in]{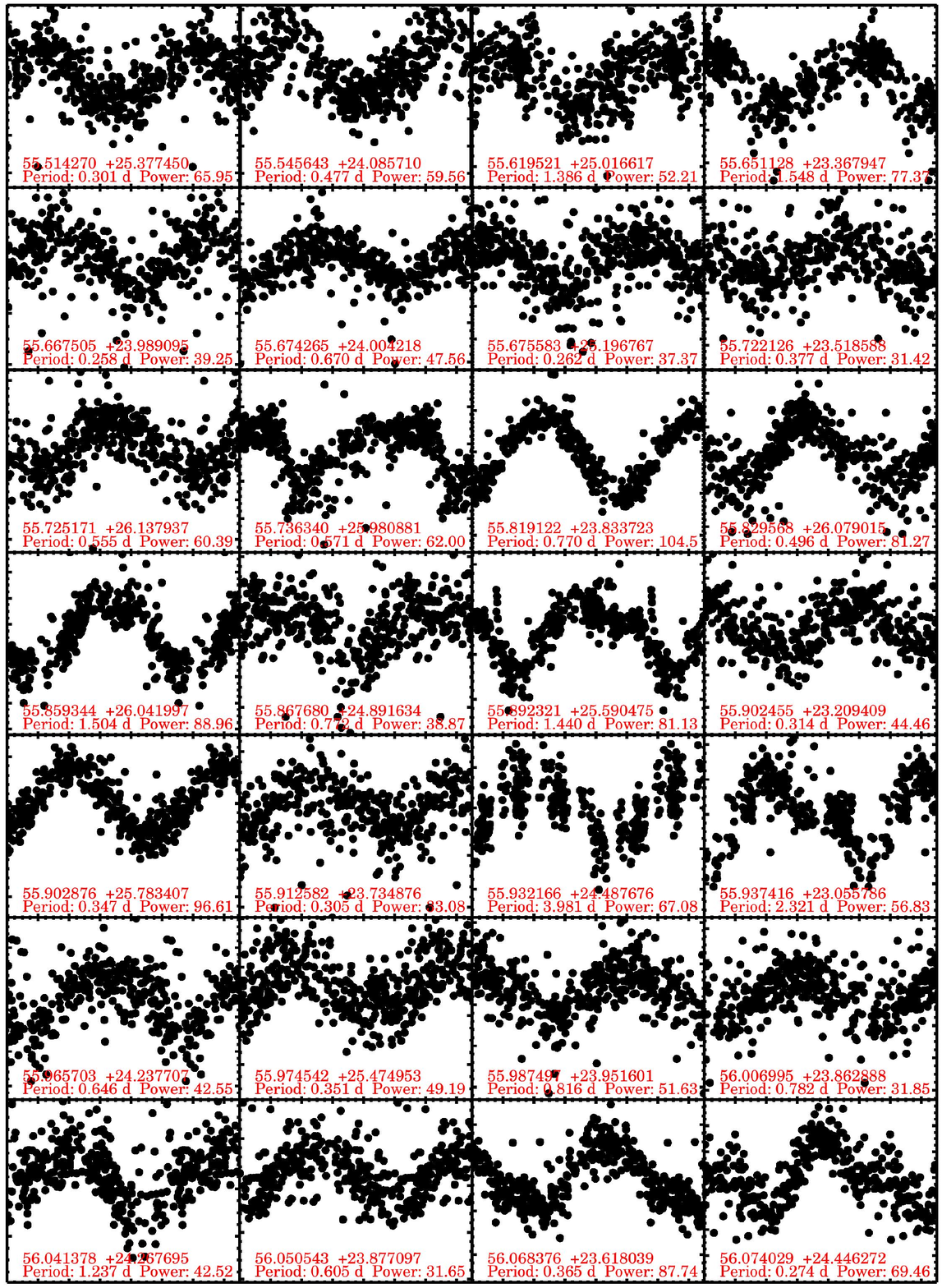}}
\caption{{Phase-folded light curves for Pleiades members with newly measured \Prot.  The axes are scaled as described in Figure~\ref{fig:periods_1}, and the RA ordering continues from Figure~\ref{fig:periods_1}. }} 
\label{fig:periods_2}
\end{figure*}

\begin{figure*}[t!] 
\centerline{\includegraphics[angle=0, width=6.5in]{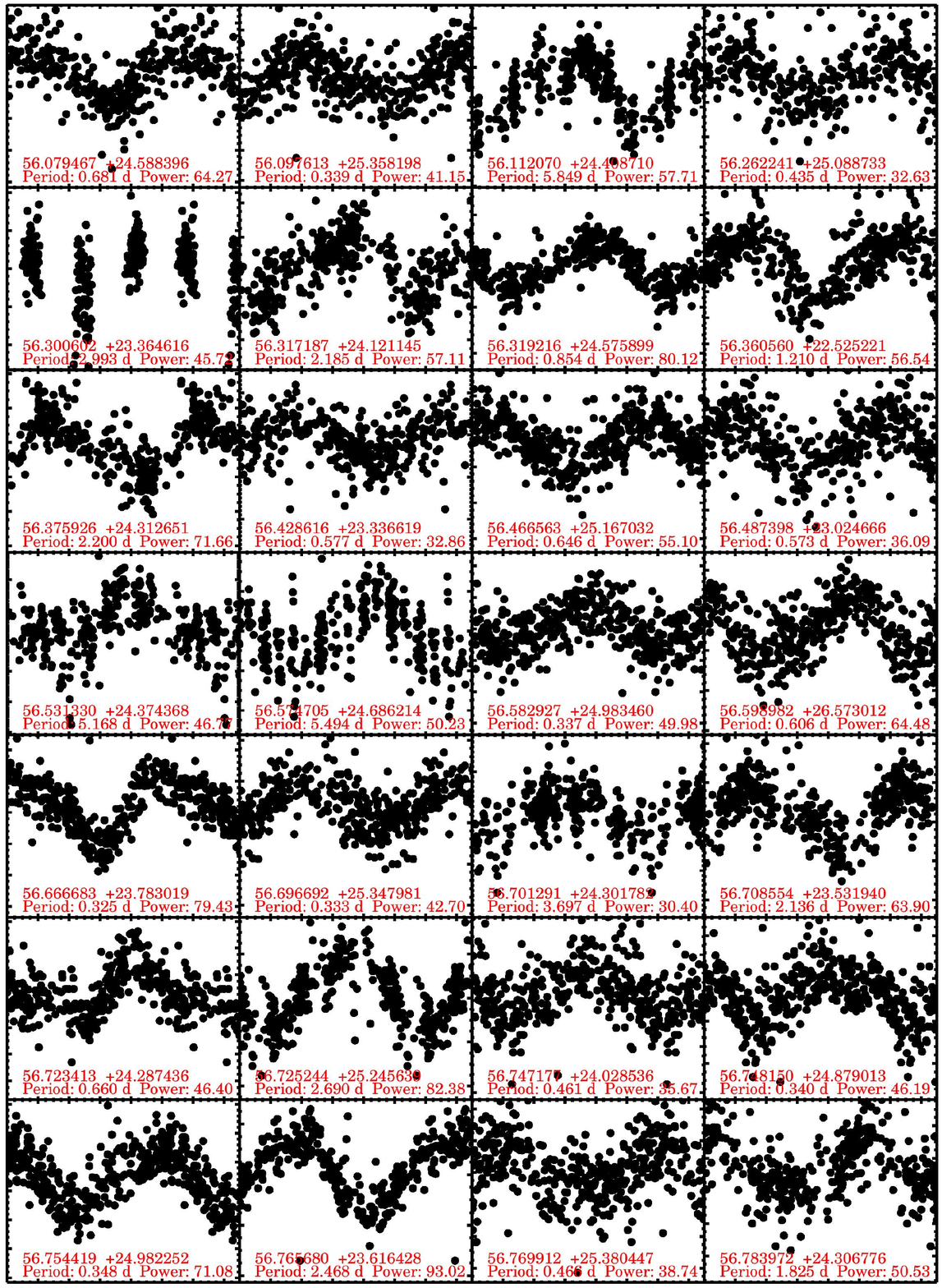}}
\caption{{Phase-folded light curves for Pleiades members with newly measured \Prot. The axes are scaled as described in Figure~\ref{fig:periods_1}, and the RA ordering continues from Figure~\ref{fig:periods_2}. } } 
\label{fig:periods_3}
\end{figure*}

\begin{figure*}[t!] 
\centerline{\includegraphics[angle=0, width=6.5in]{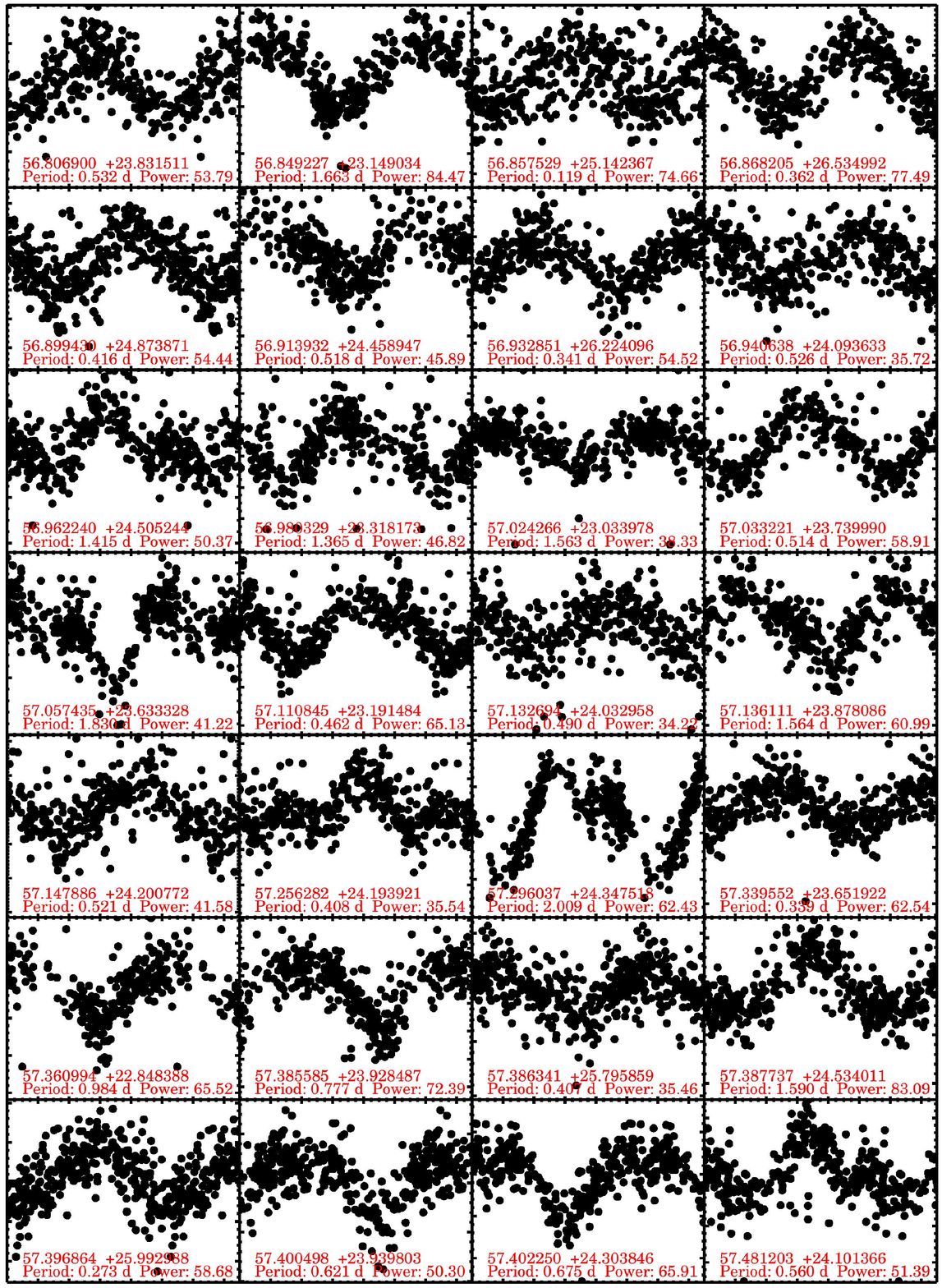}}
\caption{{Phase-folded light curves for Pleiades members with newly measured \Prot.  The axes are scaled as described in Figure~\ref{fig:periods_1}, and the RA ordering continues from Figure~\ref{fig:periods_3}.}  } 
\label{fig:periods_4}
\end{figure*}

\begin{figure*}[t!] 
\centerline{\includegraphics[angle=0, width=6.5in]{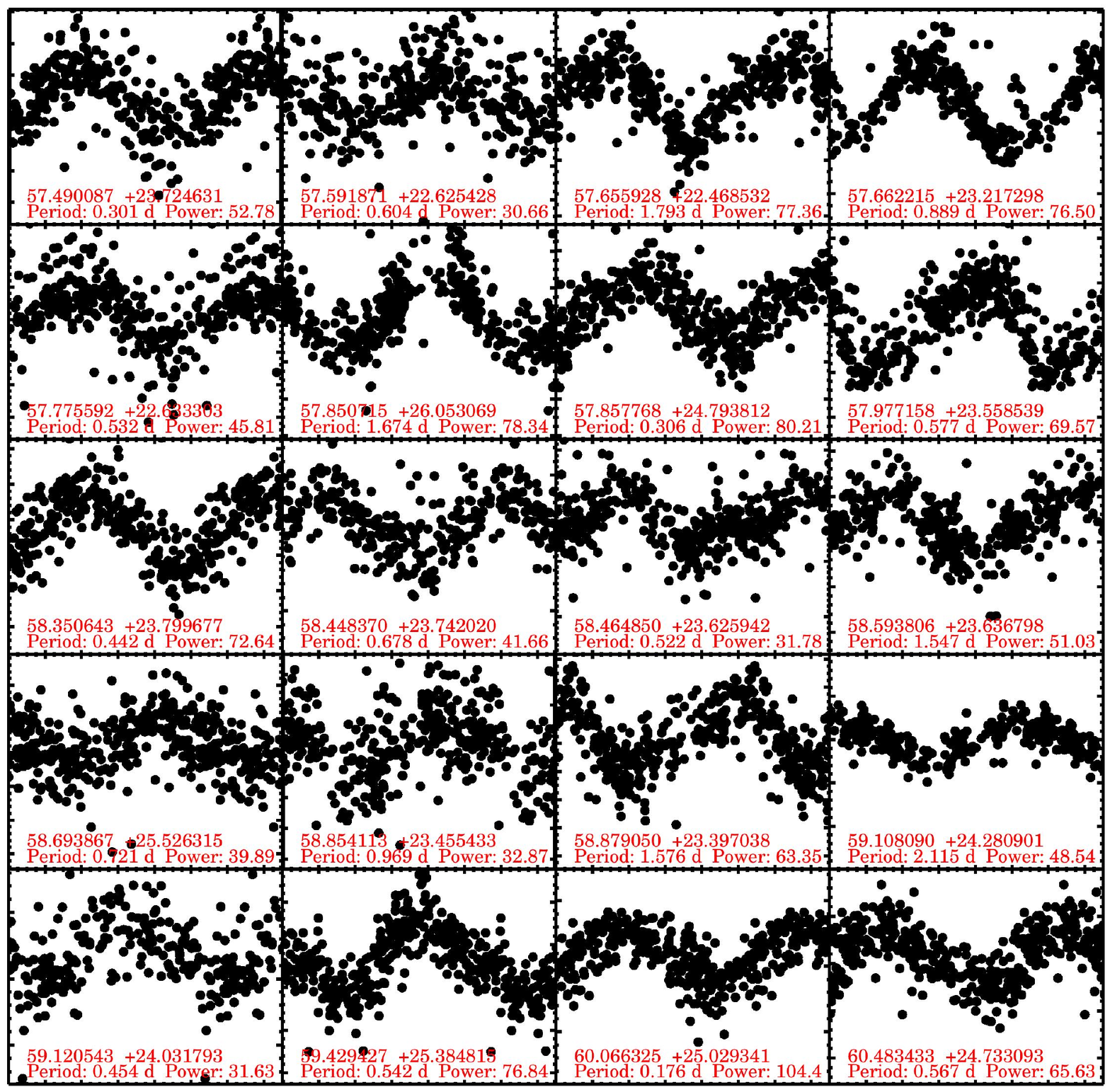}}
\caption{{Phase-folded light curves for Pleiades members with newly measured \Prot.  The axes are scaled as described in Figure~\ref{fig:periods_1}, and the RA ordering continues from Figure~\ref{fig:periods_4}. } } 
\label{fig:periods_5}
\end{figure*}


\clearpage
\LongTables
\begin{landscape}
\renewcommand{\thefootnote}{\alph{footnote}}
\begin{deluxetable}{ccccccccccccccc}
\tablewidth{0pt}
\tabletypesize{\tiny}
\tablecaption{POCS Pleiades Periods  \label{tab:Periods}}
\tablehead{
  \colhead{RA} &
  \colhead{Dec} &
  \colhead{$V$} &
  \colhead{$V$} &
  \colhead{$K$} &
  \colhead{$K$ err} &
  \colhead{$K$} &
  \colhead{$\Delta$$(V-K)$} &
  \colhead{Mass} &
  \colhead{Phot. } &
  \colhead{POCS} &
  \colhead{POCS} &
  \colhead{POCS} &
  \colhead{POCS} &
\colhead{HATnet} \\
  \colhead{(J2000)} &
  \colhead{(J2000)} &
  \colhead{(mag)} &
  \colhead{source} &
  \colhead{(mag)} &
  \colhead{(mag)} &
  \colhead{source} &
  \colhead{(mag)} &
  \colhead{(\Msun)} &
  \colhead{Bin.?} &
  \colhead{\Prot (d)} &
  \colhead{Power } &
  \colhead{Amp.} &
  \colhead{epochs} &
\colhead{\Prot\ (d)}}
\startdata
  52.49466511  &    23.37155514  &  \nodata  &  \nodata  &   11.95  &    0.0180  &    Stauffer  &  \nodata  &    0.41  &  \nodata  &   2.76  &   42.75  &     0.06  &  335  &   \nodata  \\
  52.95773498  &    22.84000116  &  \nodata  &  \nodata  &   12.75  &    0.0240  &       2MASS  &  \nodata  &    0.27  &  \nodata  &   0.57  &   53.95  &     0.08  &  336  &   \nodata  \\
  53.06352439  &    26.69864157  &   16.75  &       Kamai  &   11.73  &    0.0200  &    Stauffer  &    0.07  &    0.45  &  0  &   0.44  &   71.05  &     0.05  &  303  &   \nodata  \\
  53.14033811  &    24.49296746  &  \nodata  &  \nodata  &   12.19  &    0.0210  &    Stauffer  &  \nodata  &    0.37  &  \nodata  &   0.59  &   68.52  &     0.08  &  335  &   \nodata  \\
  53.21395850  &    25.91573260  &  \nodata  &  \nodata  &   13.00  &  \nodata  &      UKIDSS  &  \nodata  &    0.23  &  \nodata  &   0.45  &   39.90  &     0.09  &  303  &   \nodata  \\
  53.30992987  &    24.83533410  &  \nodata  &  \nodata  &   12.86  &    0.0260  &       2MASS  &  \nodata  &    0.25  &  \nodata  &   0.29  &   32.18  &     0.08  &  303  &   \nodata  \\
  53.37487945  &    26.73437085  &  \nodata  &  \nodata  &   13.26  &  \nodata  &        DANCe  &  \nodata  &    0.19  &  \nodata  &   0.16  &   31.34  &     0.11  &  303  &   \nodata  \\
  53.95691042  &    25.07983834  &   16.45  &       Kamai  &   11.69  &    0.0220  &    Stauffer  &   -0.04  &    0.45  &  0  &   1.74  &   36.05  &     0.02  &  303  &   \nodata  \\
  54.06793048  &    25.14675643  &   16.35  &       Kamai  &   11.31  &    0.0210  &    Stauffer  &    0.28  &    0.52  &  1  &   0.39  &   62.50  &     0.04  &  304  &     0.39  \\
  54.09312654  &    22.74236956  &  \nodata  &  \nodata  &   11.91  &    0.0180  &    Stauffer  &  \nodata  &    0.42  &  \nodata  &   0.67  &   41.33  &     0.04  &  339  &   \nodata  \\
  54.26437463  &    24.74277685  &   16.00  &    Stauffer  &   11.44  &    0.0160  &    Stauffer  &   -0.05  &    0.49  &  0  &   0.21  &   95.29  &     0.04  &  304  &     0.21  \\
  54.34224470  &    24.90448260  &  \nodata  &  \nodata  &   12.82  &    0.0260  &       2MASS  &  \nodata  &    0.26  &  \nodata  &   0.36  &   32.81  &     0.03  &  304  &   \nodata  \\
  54.45384442  &    26.86271968  &  \nodata  &  \nodata  &   12.59  &    0.0250  &    Stauffer  &  \nodata  &    0.30  &  \nodata  &   0.62  &   46.50  &     0.06  &  305  &   \nodata  \\
  54.50866201  &    24.33753119  &   17.60  &    Stauffer  &   12.26  &    0.0190  &    Stauffer  &   -0.06  &    0.36  &  0  &   1.31  &   54.23  &     0.06  &  334  &   \nodata  \\
  54.55440308  &    24.63795865  &   18.30  &    Stauffer  &   12.79  &    0.0190  &    Stauffer  &   -0.18  &    0.26  &  0  &   0.38  &   33.85  &     0.09  &  334  &   \nodata  \\
  54.60338648  &    26.25648344  &   17.60  &    Stauffer  &   12.22  &    0.0210  &    Stauffer  &   -0.01  &    0.36  &  0  &   1.79  &   94.16  &     0.12  &  305  &   \nodata  \\
  54.68052998  &    25.37414464  &   16.40  &    Stauffer  &   11.32  &    0.0240  &    Stauffer  &    0.30  &    0.52  &  1  &   0.50  &   41.92  &     0.03  &  305  &   \nodata  \\
  54.88470657  &    24.26698474  &  \nodata  &  \nodata  &   12.20  &    0.0240  &    Stauffer  &  \nodata  &    0.37  &  \nodata  &   0.72  &   59.47  &     0.06  &  335  &   \nodata  \\
  54.92152693  &    23.47259278  &   18.80  &    Stauffer  &   13.11  &    0.0270  &    Stauffer  &   -0.21  &    0.21  &  0  &   0.27  &   32.16  &     0.12  &  336  &   \nodata  \\
  55.02936853  &    22.64644641  &   19.20  &    Stauffer  &   13.05  &    0.0280  &    Stauffer  &    0.08  &    0.22  &  0  &   0.48  &   51.16  &     0.18  &  335  &   \nodata  \\
  55.04051866  &    23.17547150  &  \nodata  &  \nodata  &   12.65  &    0.0200  &    Stauffer  &  \nodata  &    0.29  &  \nodata  &   0.50  &   59.40  &     0.07  &  336  &   \nodata  \\
  55.12990545  &    25.14797490  &   17.17  &    Stauffer  &   11.71  &    0.0200  &    Stauffer  &    0.25  &    0.45  &  1  &   0.30  &   57.01  &     0.05  &  305  &     0.42  \\
  55.13581401  &    25.47781023  &   19.50  &    Stauffer  &   13.18  &    0.0340  &    Stauffer  &    0.13  &    0.20  &  0  &   0.55  &   41.09  &     0.09  &  305  &   \nodata  \\
  55.33281900  &    25.11356486  &   18.20  &    Stauffer  &   12.83  &    0.0250  &    Stauffer  &   -0.28  &    0.26  &  0  &   0.60  &   65.92  &     0.06  &  305  &   \nodata  \\
  55.35996882  &    23.13382276  &  \nodata  &  \nodata  &   12.62  &    0.0200  &    Stauffer  &  \nodata  &    0.29  &  \nodata  &   0.84  &   66.57  &     0.10  &  336  &   \nodata  \\
  55.46811989  &    24.69928730  &   18.90  &    Stauffer  &   13.05  &    0.0200  &    Stauffer  &   -0.09  &    0.22  &  0  &   0.35  &   51.14  &     0.11  &  334  &   \nodata  \\
  55.47588758  &    25.72952459  &   16.79  &    Stauffer  &   11.91  &    0.0210  &    Stauffer  &   -0.11  &    0.42  &  0  &   6.36  &   48.18  &     0.04  &  305  &   \nodata  \\
  55.51205208  &    24.21002278  &   16.41  &    Stauffer  &   11.61  &    0.0190  &    Stauffer  &    0.02  &    0.47  &  0  &   0.70  &   48.36  &     0.04  &  334  &     0.70  \\
  55.51427049  &    25.37745032  &   18.10  &    Stauffer  &   12.44  &    0.0260  &    Stauffer  &    0.05  &    0.32  &  0  &   0.30  &   65.95  &     0.08  &  305  &   \nodata  \\
  55.54564278  &    24.08571020  &   15.95  &    Stauffer  &   11.15  &    0.0190  &    Stauffer  &    0.21  &    0.55  &  1  &   0.48  &   59.56  &     0.03  &  334  &     0.48  \\
  55.61952053  &    25.01661724  &   16.48  &    Stauffer  &   11.67  &    0.0220  &    Stauffer  &   -0.00  &    0.46  &  0  &   1.39  &   52.21  &     0.03  &  305  &   \nodata  \\
  55.65112799  &    23.36794738  &   17.60  &    Stauffer  &   12.32  &    0.0190  &    Stauffer  &   -0.11  &    0.35  &  0  &   1.55  &   77.37  &     0.09  &  295  &   \nodata  \\
  55.66750543  &    23.98909457  &   15.72  &    Stauffer  &   11.14  &    0.0190  &    Stauffer  &    0.12  &    0.55  &  0  &   0.26  &   39.26  &     0.03  &  294  &     0.26  \\
  55.67426506  &    24.00421795  &  \nodata  &  \nodata  &   12.71  &    0.0200  &    Stauffer  &  \nodata  &    0.28  &  \nodata  &   0.67  &   47.57  &     0.07  &  294  &   \nodata  \\
  55.67558280  &    25.19676662  &   19.61  &    Stauffer  &   13.39  &    0.0380  &    Stauffer  &   -0.02  &    0.18  &  0  &   0.26  &   37.37  &     0.15  &  305  &   \nodata  \\
  55.72212582  &    23.51858757  &  \nodata  &  \nodata  &   12.21  &  \nodata  &        DANCe  &  \nodata  &    0.37  &  \nodata &   0.38  &   31.43  &     0.03  &  295  &   \nodata  \\
  55.72517121  &    26.13793696  &   18.70  &    Stauffer  &   12.82  &    0.0260  &    Stauffer  &    0.02  &    0.26  &  0  &   0.56  &   60.39  &     0.07  &  305  &   \nodata  \\
  55.73633992  &    25.98088064  &   18.00  &    Stauffer  &   11.96  &    0.0280  &    Stauffer  &    0.47  &    0.41  &  1  &   0.57  &   62.00  &     0.08  &  305  &   \nodata  \\
  55.81912223  &    23.83372277  &   18.48  &    Stauffer  &   12.40  &    0.0200  &    Stauffer  &    0.32  &    0.33  &  1  &   0.77  &  104.55  &     0.20  &  294  &   \nodata  \\
  55.82956767  &    26.07901470  &   18.00  &    Stauffer  &   12.31  &    0.0210  &    Stauffer  &    0.13  &    0.35  &  0  &   0.50  &   81.28  &     0.11  &  305  &   \nodata  \\
  55.85934378  &    26.04199654  &   16.56  &    Stauffer  &   11.89  &    0.0220  &    Stauffer  &   -0.18  &    0.42  &  0  &   1.50  &   88.97  &     0.05  &  305  &   \nodata  \\
  55.86767976  &    24.89163425  &   17.53  &    Stauffer  &   12.17  &    0.0200  &    Stauffer  &   -0.01  &    0.37  &  0  &   0.77  &   38.88  &     0.07  &  305  &   \nodata  \\
  55.89232064  &    25.59047460  &   16.99  &    Stauffer  &   12.00  &    0.0220  &    Stauffer  &   -0.12  &    0.40  &  0  &   1.44  &   81.14  &     0.09  &  305  &   \nodata  \\
  55.90245549  &    23.20940912  &   18.00  &    Stauffer  &   12.37  &    0.0200  &    Stauffer  &    0.06  &    0.33  &  0  &   0.31  &   44.46  &     0.09  &  295  &   \nodata  \\
  55.90287551  &    25.78340745  &   16.83  &    Stauffer  &   12.05  &    0.0220  &    Stauffer  &   -0.23  &    0.39  &  0  &   0.35  &   96.61  &     0.08  &  305  &   \nodata  \\
  55.91258238  &    23.73487587  &   18.05  &    Stauffer  &   12.27  &    0.0190  &    Stauffer  &    0.20  &    0.35  &  1  &   0.31  &   33.08  &     0.05  &  294  &   \nodata  \\
  55.93216646  &    24.48767634  &   16.10  &    Stauffer  &   11.50  &    0.0200  &    Stauffer  &   -0.05  &    0.48  &  0  &   3.98  &   67.09  &     0.04  &  293  &   \nodata  \\
  55.93741630  &    23.05578639  &   16.83  &    Stauffer  &   11.73  &    0.0190  &    Stauffer  &    0.09  &    0.45  &  0  &   2.32  &   56.83  &     0.05  &  293  &   \nodata  \\
  55.96570267  &    24.23770738  &  \nodata  &  \nodata  &   12.38  &    0.0210  &    Stauffer  &  \nodata  &    0.33  &  \nodata  &   0.65  &   42.55  &     0.09  &  293  &   \nodata  \\
  55.97454226  &    25.47495274  &   16.15  &    Stauffer  &   11.47  &    0.0220  &    Stauffer  &    0.00  &    0.49  &  0  &   0.35  &   49.19  &     0.02  &  305  &     0.35  \\
  55.98749666  &    23.95160059  &   17.93  &    Stauffer  &   12.30  &    0.0190  &    Stauffer  &    0.09  &    0.35  &  0  &   0.82  &   51.63  &     0.07  &  294  &   \nodata  \\
  56.00699494  &    23.86288793  &   17.22  &    Stauffer  &   11.99  &    0.0200  &    Stauffer  &   -0.01  &    0.40  &  0  &   0.78  &   31.86  &     0.03  &  294  &   \nodata  \\
  56.04137815  &    24.26769482  &   16.36  &       Kamai  &   11.53  &    0.0200  &    Stauffer  &    0.06  &    0.48  &  0  &   1.24  &   42.52  &     0.03  &  293  &     1.24  \\
  56.05054289  &    23.87709713  &   17.90  &    Stauffer  &   12.57  &    0.0190  &    Stauffer  &   -0.19  &    0.30  &  0  &   0.61  &   31.66  &     0.05  &  294  &   \nodata  \\
  56.06837622  &    23.61803853  &   17.69  &    Stauffer  &   11.89  &    0.0190  &    Stauffer  &    0.37  &    0.42  &  1  &   0.37  &   87.74  &     0.08  &  294  &   \nodata  \\
  56.07402913  &    24.44627184  &   16.62  &    Stauffer  &   11.69  &    0.0200  &    Stauffer  &    0.05  &    0.45  &  0  &   0.27  &   69.46  &     0.06  &  293  &   \nodata  \\
  56.07946744  &    24.58839573  &   17.83  &    Stauffer  &   12.44  &    0.0200  &    Stauffer  &   -0.10  &    0.32  &  0  &   0.68  &   64.27  &     0.11  &  293  &   \nodata  \\
  56.09761289  &    25.35819838  &   16.80  &    Stauffer  &   11.61  &    0.0180  &    Stauffer  &    0.19  &    0.47  &  0  &   0.34  &   41.15  &     0.03  &  305  &   \nodata  \\
  56.11206969  &    24.40870971  &   16.07  &    Stauffer  &   11.45  &    0.0190  &    Stauffer  &   -0.02  &    0.49  &  0  &   5.85  &   57.72  &     0.04  &  293  &   \nodata  \\
  56.26224139  &    25.08873294  &   18.54  &    Stauffer  &   12.85  &    0.0200  &    Stauffer  &   -0.10  &    0.25  &  0  &   0.44  &   32.63  &     0.07  &  305  &   \nodata  \\
  56.30060152  &    23.36461647  &   17.64  &    Stauffer  &   12.24  &    0.0190  &    Stauffer  &   -0.02  &    0.36  &  0  &   2.99  &   45.72  &     0.07  &  292  &   \nodata  \\
  56.31718678  &    24.12114504  &   16.00  &    Stauffer  &   11.63  &    0.0230  &    Stauffer  &   -0.24  &    0.46  &  0  &   2.19  &   57.12  &     0.05  &  298  &   \nodata  \\
  56.31921573  &    24.57589935  &   16.18  &    Stauffer  &   11.43  &    0.0190  &    Stauffer  &    0.05  &    0.50  &  0  &   0.85  &   80.13  &     0.05  &  293  &     0.86  \\
  56.36056011  &    22.52522100  &   17.40  &    Stauffer  &   12.25  &    0.0180  &    Stauffer  &   -0.16  &    0.36  &  0  &   1.21  &   56.55  &     0.15  &  297  &   \nodata  \\
  56.37592551  &    24.31265149  &   15.74  &    Stauffer  &   11.13  &    0.0190  &    Stauffer  &    0.14  &    0.55  &  0  &   2.20  &   71.66  &     0.04  &  293  &     2.19  \\
  56.42861587  &    23.33661859  &  \nodata  &  \nodata  &   13.20  &  \nodata  &      UKIDSS  &  \nodata  &    0.20  &  \nodata  &   0.58  &   32.87  &     0.13  &  292  &   \nodata  \\
  56.46656334  &    25.16703178  &   18.70  &    Stauffer  &   12.81  &    0.0220  &    Stauffer  &    0.03  &    0.26  &  0  &   0.65  &   55.11  &     0.09  &  305  &   \nodata  \\
  56.48739829  &    23.02466598  &   17.90  &    Stauffer  &   12.58  &    0.0190  &    Stauffer  &   -0.20  &    0.30  &  0  &   0.57  &   36.09  &     0.07  &  292  &   \nodata  \\
  56.53132955  &    24.37436769  &   14.71  &    Stauffer  &   10.86  &    0.0190  &    Stauffer  &   -0.04  &    0.60  &  0  &   5.17  &   46.77  &     0.02  &  293  &     5.17  \\
  56.57470506  &    24.68621385  &   16.20  &    Stauffer  &   11.69  &    0.0190  &    Stauffer  &   -0.19  &    0.45  &  0  &   5.49  &   50.23  &     0.04  &  295  &     5.49  \\
  56.58292723  &    24.98346040  &   17.40  &    Stauffer  &   11.97  &    0.0200  &    Stauffer  &    0.12  &    0.41  &  0  &   0.34  &   49.98  &     0.03  &  305  &   \nodata  \\
  56.59898192  &    26.57301170  &   18.60  &    Stauffer  &   12.84  &    0.0250  &    Stauffer  &   -0.06  &    0.26  &  0  &   0.61  &   64.49  &     0.08  &  305  &   \nodata  \\
  56.66668339  &    23.78301855  &  \nodata  &  \nodata  &   11.92  &  \nodata  &        DANCe  &  \nodata  &    0.42  &  \nodata  &   0.33  &   79.44  &     0.12  &  293  &   \nodata  \\
  56.69669244  &    25.34798089  &   18.70  &    Stauffer  &   12.53  &    0.0250  &    Stauffer  &    0.31  &    0.31  &  1  &   0.33  &   42.71  &     0.06  &  305  &   \nodata  \\
  56.70129124  &    24.30178170  &   16.24  &    Stauffer  &   11.72  &    0.0200  &    Stauffer  &   -0.20  &    0.45  &  0  &   3.70  &   30.40  &     0.01  &  295  &     3.72  \\
  56.70855377  &    23.53193973  &   17.80  &    Stauffer  &   12.39  &    0.0190  &    Stauffer  &   -0.07  &    0.33  &  0  &   2.14  &   63.91  &     0.08  &  296  &   \nodata  \\
  56.72341315  &    24.28743618  &   16.27  &    Stauffer  &   11.21  &    0.0230  &    Stauffer  &    0.32  &    0.54  &  1  &   0.66  &   46.41  &     0.04  &  295  &   \nodata  \\
  56.72524431  &    25.24563863  &   16.19  &    Stauffer  &   11.49  &    0.0200  &    Stauffer  &   -0.00  &    0.49  &  0  &   2.69  &   82.39  &     0.03  &  305  &   \nodata  \\
  56.74717714  &    24.02853609  &   17.60  &    Stauffer  &   12.11  &    0.0240  &    Stauffer  &    0.09  &    0.38  &  0  &   0.46  &   35.68  &     0.06  &  293  &   \nodata  \\
  56.74814984  &    24.87901350  &   18.40  &    Stauffer  &   12.35  &    0.0240  &    Stauffer  &    0.32  &    0.34  &  1  &   0.34  &   46.19  &     0.07  &  305  &   \nodata  \\
  56.75441943  &    24.98225183  &  \nodata  &  \nodata  &   12.34  &    0.0240  &    Stauffer  &  \nodata  &    0.34  &  \nodata  &   0.35  &   71.08  &     0.10  &  305  &   \nodata  \\
  56.76568048  &    23.61642753  &   15.35  &    Stauffer  &   10.60  &    0.0190  &    Stauffer  &    0.48  &    0.65  &  1  &   2.47  &   93.02  &     0.03  &  293  &     2.45  \\
  56.76991193  &    25.38044651  &   16.35  &    Stauffer  &   11.26  &    0.0180  &    Stauffer  &    0.32  &    0.53  &  1  &   0.47  &   38.75  &     0.03  &  305  &   \nodata  \\
  56.78397188  &    24.30677566  &   16.95  &    Stauffer  &   11.90  &    0.0190  &    Stauffer  &   -0.04  &    0.42  &  0  &   1.83  &   50.53  &     0.04  &  295  &   \nodata  \\
  56.80690016  &    23.83151104  &   15.78  &    Stauffer  &   11.09  &    0.0190  &    Stauffer  &    0.20  &    0.56  &  1  &   0.53  &   53.79  &     0.03  &  293  &   \nodata  \\
  56.84922746  &    23.14903368  &   17.33  &    Stauffer  &   12.07  &    0.0180  &    Stauffer  &   -0.03  &    0.39  &  0  &   1.66  &   84.48  &     0.08  &  296  &   \nodata  \\
  56.85752870  &    25.14236677  &   16.91  &    Stauffer  &   11.45  &    0.0180  &    Stauffer  &    0.40  &    0.49  &  1  &   0.12  &   74.67  &     0.09  &  304  &     0.14  \\
  56.86820506  &    26.53499222  &  \nodata  &  \nodata  &   12.39  &    0.0190  &    Stauffer  &  \nodata  &    0.33  &  \nodata  &   0.36  &   77.49  &     0.13  &  303  &   \nodata  \\
  56.89942978  &    24.87387056  &   18.78  &    Stauffer  &   12.84  &    0.0240  &    Stauffer  &    0.05  &    0.26  &  0  &   0.42  &   54.44  &     0.10  &  304  &   \nodata  \\
  56.91393165  &    24.45894716  &   17.56  &    Stauffer  &   12.16  &    0.0240  &    Stauffer  &    0.02  &    0.37  &  0  &   0.52  &   45.89  &     0.08  &  293  &   \nodata  \\
  56.93285137  &    26.22409550  &  \nodata  &  \nodata  &   12.82  &    0.0290  &    Stauffer  &  \nodata  &    0.26  &  \nodata  &   0.34  &   54.53  &     0.09  &  302  &   \nodata  \\
  56.94063826  &    24.09363326  &  \nodata  &  \nodata  &   12.35  &  \nodata  &        DANCe  &  \nodata  &    0.34  &  \nodata  &   0.53  &   35.73  &     0.07  &  291  &   \nodata  \\
  56.96224034  &    24.50524354  &   16.13  &    Stauffer  &   11.47  &    0.0230  &    Stauffer  &   -0.01  &    0.49  &  0  &   1.42  &   50.37  &     0.03  &  293  &     1.41  \\
  56.98032942  &    23.31817262  &   17.33  &    Stauffer  &   12.15  &    0.0190  &    Stauffer  &   -0.10  &    0.38  &  0  &   1.37  &   46.82  &     0.04  &  292  &   \nodata  \\
  57.02426564  &    23.03397771  &   16.16  &    Stauffer  &   11.55  &    0.0190  &    Stauffer  &   -0.08  &    0.48  &  0  &   1.56  &   38.34  &     0.03  &  292  &     1.56  \\
  57.03322114  &    23.73998951  &   17.30  &    Stauffer  &   12.18  &    0.0190  &    Stauffer  &   -0.15  &    0.37  &  0  &   0.51  &   58.92  &     0.07  &  291  &   \nodata  \\
  57.05743494  &    23.63332766  &   17.72  &    Stauffer  &   12.11  &    0.0180  &    Stauffer  &    0.17  &    0.38  &  1  &   1.83  &   41.22  &     0.08  &  291  &   \nodata  \\
  57.11084501  &    23.19148426  &   17.87  &    Stauffer  &   11.36  &    0.0200  &    Stauffer  &    1.00  &    0.51  &  1  &   0.46  &   65.14  &     0.07  &  292  &   \nodata  \\
  57.13269373  &    24.03295810  &   18.30  &    Stauffer  &   12.85  &    0.0210  &    Stauffer  &   -0.24  &    0.25  &  0  &   0.49  &   34.22  &     0.07  &  291  &   \nodata  \\
  57.13611052  &    23.87808619  &  \nodata  &  \nodata  &   12.87  &    0.0270  &       2MASS  &  \nodata  &    0.25  &  \nodata  &   1.56  &   61.00  &     0.11  &  291  &   \nodata  \\
  57.14788550  &    24.20077232  &   19.67  &    Stauffer  &   13.21  &    0.0210  &    Stauffer  &    0.20  &    0.20  &  1  &   0.52  &   41.58  &     0.17  &  293  &   \nodata  \\
  57.25628182  &    24.19392074  &   17.70  &    Stauffer  &   12.51  &    0.0190  &    Stauffer  &   -0.25  &    0.31  &  0  &   0.41  &   35.54  &     0.07  &  292  &   \nodata  \\
  57.29603697  &    24.34751792  &   17.43  &    Stauffer  &   12.18  &    0.0240  &    Stauffer  &   -0.07  &    0.37  &  0  &   2.01  &   62.43  &     0.11  &  292  &   \nodata  \\
  57.33955247  &    23.65192192  &   17.15  &    Stauffer  &   11.98  &    0.0200  &    Stauffer  &   -0.03  &    0.41  &  0  &   0.34  &   62.54  &     0.04  &  291  &   \nodata  \\
  57.36099383  &    22.84838830  &   18.23  &    Stauffer  &   12.51  &    0.0230  &    Stauffer  &    0.06  &    0.31  &  0  &   0.98  &   65.52  &     0.08  &  292  &   \nodata  \\
  57.38558548  &    23.92848672  &   17.91  &    Stauffer  &   12.44  &    0.0190  &    Stauffer  &   -0.05  &    0.32  &  0  &   0.78  &   72.39  &     0.12  &  291  &   \nodata  \\
  57.38634076  &    25.79585908  &   18.30  &    Stauffer  &   12.62  &    0.0220  &    Stauffer  &   -0.01  &    0.29  &  0  &   0.41  &   35.46  &     0.05  &  301  &   \nodata  \\
  57.38773738  &    24.53401103  &   16.58  &    Stauffer  &   11.74  &    0.0200  &    Stauffer  &   -0.02  &    0.45  &  0  &   1.59  &   83.09  &     0.07  &  292  &     1.59  \\
  57.39686426  &    25.99298827  &   18.30  &    Stauffer  &   12.69  &    0.0250  &    Stauffer  &   -0.08  &    0.28  &  0  &   0.27  &   58.69  &     0.08  &  302  &   \nodata  \\
  57.40049832  &    23.93980292  &   17.46  &    Stauffer  &   12.28  &    0.0190  &    Stauffer  &   -0.15  &    0.35  &  0  &   0.62  &   50.31  &     0.08  &  291  &   \nodata  \\
  57.40225000  &    24.30384552  &   17.23  &    Stauffer  &   12.43  &    0.0240  &    Stauffer  &   -0.44  &    0.32  &  0  &   0.68  &   65.92  &     0.09  &  292  &   \nodata  \\
  57.48120347  &    24.10136598  &  \nodata  &  \nodata  &   12.63  &    0.0240  &    Stauffer  &  \nodata  &    0.29  &  \nodata  &   0.56  &   51.39  &     0.10  &  291  &   \nodata  \\
  57.49008659  &    23.72463071  &  \nodata  &  \nodata  &   13.39  &    0.0220  &    Stauffer  &  \nodata  &    0.18  &  \nodata &   0.30  &   52.78  &     0.24  &  291  &   \nodata  \\
  57.59187149  &    22.62542831  &   18.09  &    Stauffer  &   12.30  &    0.0210  &    Stauffer  &    0.19  &    0.35  &  1  &   0.60  &   30.67  &     0.04  &  295  &   \nodata  \\
  57.65592827  &    22.46853150  &  \nodata  &  \nodata  &   12.37  &    0.0190  &    Stauffer  &  \nodata  &    0.34  &  \nodata &   1.79  &   77.37  &     0.08  &  295  &   \nodata  \\
  57.66221496  &    23.21729769  &   17.12  &    Stauffer  &   11.75  &    0.0190  &    Stauffer  &    0.19  &    0.44  &  1  &   0.89  &   76.51  &     0.06  &  294  &   \nodata  \\
  57.77559169  &    22.63330340  &  \nodata  &  \nodata  &   12.98  &    0.0290  &    Stauffer  &  \nodata  &    0.23  &  \nodata &   0.53  &   45.82  &     0.10  &  295  &   \nodata  \\
  57.85071482  &    26.05306881  &   16.68  &    Stauffer  &   11.68  &    0.0230  &    Stauffer  &    0.08  &    0.46  &  0  &   1.67  &   78.35  &     0.05  &  301  &     1.67  \\
  57.85776837  &    24.79381174  &   15.83  &    Stauffer  &   11.19  &    0.0190  &    Stauffer  &    0.12  &    0.54  &  0  &   0.31  &   80.22  &     0.04  &  305  &     0.31  \\
  57.97715824  &    23.55853930  &   17.53  &    Stauffer  &   11.78  &    0.0160  &    Stauffer  &    0.38  &    0.44  &  1  &   0.58  &   69.58  &     0.09  &  295  &   \nodata  \\
  58.35064263  &    23.79967662  &   18.10  &    Stauffer  &   12.55  &    0.0220  &    Stauffer  &   -0.06  &    0.30  &  0  &   0.44  &   72.65  &     0.11  &  292  &   \nodata  \\
  58.44836996  &    23.74201991  &   17.64  &    Stauffer  &   12.28  &    0.0200  &    Stauffer  &   -0.05  &    0.35  &  0  &   0.68  &   41.66  &     0.06  &  292  &   \nodata  \\
  58.46485025  &    23.62594161  &  \nodata  &  \nodata  &   13.10  &  \nodata  &      UKIDSS  &  \nodata  &    0.21  &  \nodata &   0.52  &   31.79  &     0.12  &  292  &   \nodata  \\
  58.59380575  &    23.63679846  &   18.00  &    Stauffer  &   12.57  &    0.0230  &    Stauffer  &   -0.14  &    0.30  &  0  &   1.55  &   51.03  &     0.08  &  291  &   \nodata  \\
  58.69386732  &    25.52631461  &  \nodata  &  \nodata  &   12.68  &    0.0260  &    Stauffer  &  \nodata  &    0.28  &  \nodata &   0.72  &   39.89  &     0.08  &  304  &   \nodata  \\
  58.85411337  &    23.45543324  &  \nodata  &  \nodata  &   12.59  &    0.0250  &       2MASS  &  \nodata  &    0.30  &  \nodata  &   0.97  &   32.87  &     0.02  &  293  &   \nodata  \\
  58.87904998  &    23.39703776  &   17.21  &    Stauffer  &   12.18  &    0.0200  &    Stauffer  &   -0.20  &    0.37  &  0  &   1.58  &   63.35  &     0.05  &  293  &   \nodata  \\
  59.10809025  &    24.28090082  &   15.03  &    Stauffer  &   11.01  &    0.0180  &    Stauffer  &   -0.08  &    0.57  &  0  &   2.12  &   48.54  &     0.01  &  292  &   \nodata  \\
  59.12054309  &    24.03179251  &   17.90  &    Stauffer  &   12.64  &    0.0240  &    Stauffer  &   -0.26  &    0.29  &  0  &   0.45  &   31.63  &     0.05  &  292  &   \nodata  \\
  59.42942675  &    25.38481451  &  \nodata  &  \nodata  &   12.57  &    0.0270  &    Stauffer  &  \nodata  &    0.30  & \nodata  &   0.54  &   76.84  &     0.10  &  304  &   \nodata  \\
  60.06632455  &    25.02934101  &  \nodata  &  \nodata  &   11.69  &    0.0180  &       2MASS  &  \nodata  &    0.45  & \nodata  &   0.18  &  104.43  &     0.05  &  303  &   \nodata  \\
  60.48343341  &    24.73309279  &  \nodata  &  \nodata  &   12.35  &    0.0240  &       2MASS  &  \nodata  &    0.34  &  \nodata &   0.57  &   65.63  &     0.07  &  304  &   \nodata 
\enddata
\end{deluxetable}
\clearpage
\end{landscape}

\clearpage
{
\begin{deluxetable*}{ccccccccc}
\tablewidth{0pt}
\tablecaption{Periods For Field Stars Observed During PTF Pleiades Campaign  \label{tab:appendix}}
\tablehead{
  \colhead{RA} &
  \colhead{Dec} &
  \colhead{$R_{PTF}$} &
  \colhead{$R_{PTF}$ err} &
  \colhead{POCS} &
  \colhead{POCS} &
  \colhead{POCS} &
  \colhead{POCS} & 
  \colhead{Notes} \\
  \colhead{(J2000)} &
  \colhead{(J2000)} &
  \colhead{(mag)} &
  \colhead{(mag)} &
  \colhead{\Prot} &
  \colhead{Power } &
  \colhead{Amp.} &
  \colhead{epochs} &
  \colhead{} }
\startdata
51.94879326	&	24.24068055	&	15.24	&	0.016	&	0.71	&	57.74	&	0.04	&	337	&		\\
52.00883989	&	23.53802349	&	17.69	&	0.090	&	0.12	&	53.50	&	0.20	&	335	&		\\
52.02393778	&	23.12858749	&	18.16	&	0.130	&	0.14	&	68.41	&	0.28	&	335	&		\\
52.92060368	&	24.35219393	&	15.56	&	0.020	&	0.67	&	61.25	&	0.06	&	335	&		\\
53.04009792	&	26.70932925	&	16.32	&	0.030	&	0.67	&	61.70	&	0.08	&	303	&		\\
53.06836191	&	26.71902482	&	18.01	&	0.160	&	0.37	&	99.67	&	0.39	&	303	&		\\
53.09808043	&	26.73725839	&	15.50	&	0.024	&	1.95	&	83.11	&	0.07	&	303	& HAT 214-16653\tablenotemark{a} \\
53.16238823	&	26.42155890	&	14.67	&	0.010	&	7.18	&	91.94	&	0.04	&	303	&		\\
53.24573261	&	24.08963059	&	18.37	&	0.120	&	0.28	&	53.69	&	0.28	&	337	&		\\
53.30451868	&	23.99756412	&	15.82	&	0.030	&	1.70	&	71.59	&	0.08	&	337	&		\\
53.32414716	&	25.63574174	&	14.43	&	0.021	&	0.41	&	65.55	&	0.05	&	304	&		\\
53.32645996	&	25.96274168	&	17.24	&	0.050	&	0.12	&	53.93	&	0.09	&	303	&		\\
53.46385766	&	24.21337868	&	17.05	&	0.040	&	0.13	&	69.67	&	0.11	&	336	&		\\
53.48379170	&	25.67595397	&	17.75	&	0.065	&	0.58	&	48.94	&	0.11	&	304	&		\\
53.49683966	&	24.13472687	&	17.70	&	0.155	&	0.20	&	76.66	&	0.35	&	335	&		\\
53.50491740	&	25.36498503	&	16.82	&	0.060	&	0.28	&	72.27	&	0.12	&	304	&		\\
53.51194633	&	25.65112676	&	17.27	&	0.090	&	0.17	&	69.48	&	0.21	&	304	&		\\
53.55047644	&	26.87167932	&	18.69	&	0.150	&	0.13	&	61.11	&	0.34	&	303	&		\\
53.60923875	&	25.62690743	&	14.61	&	0.020	&	0.20	&	111.63	&	0.04	&	304	&		\\
53.63960226	&	24.14289527	&	15.64	&	0.020	&	0.22	&	74.42	&	0.05	&	335	&		\\
53.65023728	&	26.81475491	&	17.81	&	0.200	&	0.15	&	89.75	&	0.50	&	304	&		\\
53.67593849	&	25.90575834	&	16.50	&	0.030	&	0.78	&	59.30	&	0.08	&	304	&		\\
53.68094857	&	22.48026418	&	15.51	&	0.018	&	1.03	&	70.41	&	0.04	&	335	&		\\
53.69424026	&	26.22319633	&	15.07	&	0.011	&	1.48	&	57.74	&	0.02	&	304	&		\\
53.77350467	&	26.08222880	&	15.77	&	0.025	&	1.08	&	44.19	&	0.07	&	304	&		\\
53.83466983	&	25.17530613	&	17.60	&	0.091	&	0.45	&	75.93	&	0.22	&	303	&		\\
53.93929183	&	25.70422374	&	16.70	&	0.060	&	0.14	&	36.24	&	0.13	&	303	&		\\
54.19947418	&	23.51579144	&	15.33	&	0.010	&	0.26	&	45.32	&	0.02	&	338	&		\\
54.33276246	&	26.19422870	&	15.28	&	0.020	&	0.21	&	93.75	&	0.05	&	305	&		\\
54.41420020	&	22.97074248	&	19.66	&	0.646	&	0.14	&	63.02	&	1.66	&	337	&		\\
54.42398191	&	23.26708362	&	15.83	&	0.020	&	0.13	&	71.68	&	0.05	&	338	&		\\
54.47309794	&	26.52419033	&	15.98	&	0.016	&	0.78	&	48.85	&	0.03	&	305	&		\\
54.54425672	&	24.14586696	&	17.36	&	0.080	&	0.13	&	41.54	&	0.15	&	335	&		\\
54.70270854	&	22.87224373	&	17.48	&	0.064	&	0.49	&	45.62	&	0.13	&	336	&		\\
54.71701864	&	25.22733869	&	15.44	&	0.017	&	0.99	&	79.50	&	0.07	&	305	&		\\
54.74292701	&	24.51279467	&	16.69	&	0.040	&	0.13	&	72.44	&	0.11	&	334	&		\\
54.76559838	&	25.71999163	&	16.77	&	0.070	&	0.13	&	89.70	&	0.18	&	305	&		\\
54.83170241	&	26.56893551	&	16.42	&	0.041	&	0.70	&	68.82	&	0.10	&	305	&		\\
54.91657927	&	24.99429467	&	16.46	&	0.053	&	0.17	&	101.78	&	0.13	&	305	&		\\
54.96611859	&	25.01020565	&	18.59	&	0.110	&	0.36	&	69.60	&	0.28	&	305	&		\\
55.06090612	&	22.96704754	&	15.52	&	0.030	&	0.64	&	79.15	&	0.06	&	335	&		\\
55.28896875	&	22.50216499	&	17.54	&	0.177	&	0.11	&	88.91	&	0.48	&	335	&		\\
55.43015735	&	25.45481442	&	19.91	&	0.575	&	0.57	&	39.48	&	1.21	&	305	&		\\
55.63105607	&	25.57347007	&	16.33	&	0.036	&	0.33	&	64.40	&	0.10	&	305	&		\\
55.63175109	&	26.80435653	&	20.22	&	0.732	&	0.13	&	40.23	&	1.91	&	305	&		\\
55.63893441	&	23.11859443	&	15.34	&	0.020	&	0.45	&	83.92	&	0.04	&	334	&		\\
55.69099572	&	22.81896800	&	13.37	&	0.008	&	7.20	&	38.74	&	0.02	&	296	& HAT 259-07524\tablenotemark{a} \\
55.70723171	&	25.55042185	&	14.20	&	0.009	&	0.11	&	50.54	&	0.02	&	305	&		\\
55.75994836	&	25.50616517	&	18.97	&	0.431	&	0.30	&	83.49	&	1.06	&	305	&		\\
55.78017367	&	26.96653374	&	16.70	&	0.102	&	0.36	&	70.90	&	0.24	&	305	& eclipsing system? \\
55.78048863	&	23.12129595	&	18.66	&	0.310	&	0.54	&	65.46	&	0.75	&	295	&		\\
55.82047107	&	23.84000817	&	18.36	&	0.213	&	0.59	&	71.78	&	0.60	&	294	&		\\
55.91060411	&	26.78811427	&	19.91	&	0.783	&	0.13	&	52.49	&	1.85	&	305	&		\\
55.93421048	&	24.50778814	&	15.02	&	0.017	&	0.17	&	80.59	&	0.04	&	293	&		\\
55.97155762	&	23.17001070	&	17.13	&	0.046	&	0.10	&	38.94	&	0.08	&	293	&		\\
55.98405336	&	24.87148133	&	16.28	&	0.070	&	0.13	&	120.49	&	0.19	&	305	&		\\
56.09297988	&	26.13287932	&	15.43	&	0.021	&	2.65	&	85.90	&	0.05	&	305	&		\\
56.22457089	&	22.83150805	&	17.29	&	0.153	&	0.13	&	78.60	&	0.34	&	296	&		\\
56.31226993	&	23.01847327	&	18.64	&	0.320	&	0.20	&	58.85	&	0.73	&	292	&		\\
56.34840165	&	25.18516272	&	16.47	&	0.030	&	0.78	&	81.83	&	0.07	&	305	&		\\
56.48350681	&	26.53822482	&	17.67	&	0.151	&	0.16	&	80.98	&	0.36	&	305	&		\\
56.49159078	&	22.47860791	&	16.65	&	0.050	&	0.16	&	70.49	&	0.12	&	297	&		\\
56.49195399	&	24.77451636	&	16.55	&	0.071	&	0.15	&	107.98	&	0.19	&	305	&		\\
56.54708239	&	22.85455220	&	16.68	&	0.059	&	0.11	&	98.75	&	0.15	&	297	&		\\
56.60629379	&	22.52109131	&	15.40	&	0.030	&	0.21	&	74.71	&	0.06	&	294	&		\\
56.60903906	&	23.83696489	&	17.80	&	0.169	&	0.61	&	71.34	&	0.37	&	293	&		\\
56.63370336	&	25.77739441	&	17.62	&	0.270	&	0.12	&	84.34	&	0.64	&	305	&		\\
56.66720415	&	23.73691040	&	16.94	&	0.076	&	0.67	&	70.87	&	0.15	&	293	&		\\
56.78816577	&	22.50875452	&	15.50	&	0.025	&	2.38	&	91.99	&	0.07	&	294	& HAT 260-17311\tablenotemark{a} \\
56.95470385	&	23.31837208	&	13.93	&	0.010	&	8.41	&	42.05	&	0.01	&	292	& HAT 260-15511\tablenotemark{a} \\
57.02932992	&	24.45347207	&	13.80	&	0.004	&	0.55	&	65.90	&	0.01	&	293	&		\\
57.27038431	&	23.49428226	&	17.17	&	0.117	&	0.16	&	87.16	&	0.28	&	293	&		\\
57.28316201	&	24.66621172	&	17.16	&	0.050	&	1.14	&	62.35	&	0.11	&	292	&		\\
57.34771579	&	26.46287240	&	17.22	&	0.117	&	0.18	&	85.55	&	0.27	&	304	&		\\
57.50125009	&	26.23309952	&	16.06	&	0.070	&	0.17	&	90.85	&	0.16	&	303	&		\\
57.52071549	&	24.84965532	&	15.77	&	0.025	&	0.86	&	50.32	&	0.06	&	302	&		\\
57.63369801	&	22.93461472	&	18.37	&	0.120	&	0.11	&	41.85	&	0.23	&	295	&		\\
57.65023870	&	24.80087266	&	13.29	&	0.006	&	7.61	&	50.46	&	0.01	&	302	&		\\
57.65520164	&	23.24005508	&	15.96	&	0.020	&	3.46	&	66.60	&	0.07	&	294	&		\\
57.79423753	&	22.85162850	&	13.94	&	0.006	&	1.40	&	54.75	&	0.02	&	295	& HAT 260-17647\tablenotemark{a} \\
57.95069142	&	23.28593777	&	15.02	&	0.010	&	0.98	&	58.74	&	0.02	&	295	& LP 357-25 \\
57.97951864	&	22.56919993	&	18.92	&	0.203	&	0.15	&	55.36	&	0.47	&	295	&		\\
58.15605533	&	26.33587884	&	17.21	&	0.124	&	0.12	&	85.49	&	0.31	&	304	&		\\
58.18445476	&	26.16550063	&	14.27	&	0.008	&	0.81	&	61.99	&	0.02	&	304	&		\\
58.23250991	&	24.43780141	&	18.42	&	0.122	&	0.77	&	62.33	&	0.31	&	293	& CSS J035255.8+242615\tablenotemark{b}	\\
58.29638364	&	23.48580383	&	15.41	&	0.030	&	0.22	&	108.64	&	0.08	&	296	&		\\
58.41222669	&	25.73144733	&	17.39	&	0.110	&	0.16	&	67.98	&	0.24	&	304	&		\\
58.47022025	&	23.44487631	&	16.04	&	0.028	&	0.85	&	97.70	&	0.08	&	296	&		\\
58.48019506	&	24.63153754	&	18.67	&	0.110	&	0.21	&	42.51	&	0.23	&	293	&		\\
58.51391773	&	23.86491914	&	15.68	&	0.019	&	7.54	&	76.75	&	0.05	&	291	&		\\
58.61281679	&	26.41789809	&	16.65	&	0.030	&	0.10	&	46.14	&	0.06	&	304	&		\\
58.61606076	&	24.27806186	&	18.66	&	0.212	&	0.55	&	58.34	&	0.57	&	292	& CSS J035427.8+241640\tablenotemark{b}	\\
58.62379156	&	25.01125225	&	17.30	&	0.100	&	0.13	&	77.92	&	0.24	&	304	&		\\
58.65999363	&	26.26945856	&	17.00	&	0.249	&	0.13	&	63.94	&	0.54	&	304	&		\\
58.71177661	&	24.06327876	&	16.42	&	0.050	&	0.17	&	83.42	&	0.11	&	291	&		\\
58.87438235	&	24.48944040	&	20.26	&	0.647	&	2.11	&	48.46	&	1.90	&	292	&		\\
58.92479274	&	22.62798755	&	16.10	&	0.028	&	0.85	&	83.53	&	0.07	&	295	&		\\
58.93820123	&	25.26666618	&	17.11	&	0.073	&	0.16	&	50.99	&	0.18	&	303	&		\\
58.96314646	&	22.89492339	&	17.70	&	0.189	&	0.60	&	68.37	&	0.48	&	295	& CSS J035551.1+225341\tablenotemark{b}	\\
58.97522679	&	23.06746821	&	19.45	&	0.289	&	0.53	&	46.92	&	0.66	&	293	& CSS J035554.0+230403\tablenotemark{b}	\\
59.03207308	&	22.90516749	&	19.03	&	0.180	&	0.33	&	56.98	&	0.36	&	295	&		\\
59.09405239	&	23.18376714	&	18.32	&	0.141	&	0.10	&	54.00	&	0.29	&	293	&		\\
59.17446697	&	22.66218246	&	18.07	&	0.192	&	0.18	&	81.20	&	0.44	&	296	&		\\
59.19322655	&	23.97335364	&	19.03	&	0.246	&	0.56	&	37.94	&	0.48	&	292	& CSS J035646.3+235824\tablenotemark{b}	\\
59.21960568	&	23.40598291	&	17.21	&	0.039	&	1.74	&	58.04	&	0.09	&	294	&		\\
59.22448748	&	23.20549547	&	15.23	&	0.008	&	0.37	&	44.74	&	0.02	&	294	&		\\
59.30690890	&	22.94937108	&	17.97	&	0.065	&	0.25	&	41.32	&	0.15	&	296	&		\\
59.58672103	&	25.33420027	&	15.96	&	0.022	&	9.98	&	43.22	&	0.08	&	304	&		\\
59.60070147	&	26.69279303	&	14.68	&	0.010	&	0.42	&	65.18	&	0.02	&	303	&		\\
59.65776395	&	25.70938149	&	16.33	&	0.070	&	0.15	&	107.01	&	0.19	&	304	&		\\
59.73946412	&	25.53273386	&	16.99	&	0.080	&	0.21	&	87.24	&	0.18	&	304	&		\\
59.75175618	&	26.11063316	&	17.44	&	0.096	&	0.13	&	91.11	&	0.26	&	303	&		\\
59.78840316	&	25.15999718	&	15.92	&	0.064	&	0.17	&	81.69	&	0.15	&	305	&		\\
60.19923012	&	25.63211714	&	17.71	&	0.110	&	0.16	&	76.53	&	0.27	&	304	&		\\
60.28371183	&	24.82612949	&	18.39	&	0.310	&	0.18	&	80.25	&	0.70	&	303	&		\\
60.59665635	&	25.75287068	&	16.19	&	0.102	&	0.16	&	112.14	&	0.26	&	304	&		\\
60.61352009	&	26.07631206	&	17.53	&	0.168	&	0.20	&	81.21	&	0.40	&	302	&		\\
60.62548353	&	25.32855704	&	17.79	&	0.160	&	0.11	&	70.67	&	0.36	&	304	&		\\
60.64253195	&	26.26916643	&	17.14	&	0.060	&	0.16	&	64.95	&	0.12	&	302	&		
\enddata
\tablenotetext{a}{These are non-members with \Prot\ measured by \citet{Hartman2010}. The \Prot\ derived here and in that work agree to better than 1\% for all these stars. }
\tablenotetext{b}{Known RR Lyrae.}
\end{deluxetable*}
}

\end{document}